\documentclass[aps,prx,twocolumn,showpacs,superscriptaddress,floatfix
,nofootinbib]{revtex4-1}
\usepackage{amsmath}
\usepackage{graphicx}
\usepackage{dcolumn}
\usepackage{bm}
\usepackage[colorlinks,citecolor=blue,linkcolor=blue,anchorcolor=blue,filecolor=blue,
urlcolor=blue]{hyperref}
\usepackage{hyperref}
\usepackage{ulem}
\usepackage{xcolor}
\usepackage[utf8]{inputenc}
\usepackage{tabularx}
\usepackage{color}
\usepackage{cellspace}
\usepackage{lipsum, babel}
\usepackage{lipsum}
\usepackage{multirow}
\usepackage{booktabs} 
\definecolor{customviolet}{RGB}{148,0,211}
\usepackage{orcidlink}
\usepackage{float}
\usepackage{adjustbox}
\newcommand{\be}{\begin{equation}}
\newcommand{\ee}{\end{equation}}
\newcommand{\bea}{\begin{eqnarray}}
\newcommand{\eea}{\end{eqnarray}}

\usepackage{orcidlink}

\begin{document}

\author{Adamu Issifu \orcidlink{0000-0002-2843-835X}} 
\email{ai@academico.ufpb.br}
\affiliation{Departamento de F\'isica e Laborat\'orio de Computa\c c\~ao Cient\'ifica Avan\c cada e Modelamento (Lab-CCAM), Instituto Tecnol\'ogico de Aeron\'autica, DCTA, 12228-900, S\~ao Jos\'e dos Campos, SP, Brazil} 

\author{Franciele M. da Silva \orcidlink{0000-0003-2568-2901}} 
\email{franciele.m.s@ufsc.br}

\affiliation{Departamento de F\'isica, CFM - Universidade Federal de Santa Catarina; \\ C.P. 476, CEP 88.040-900, Florian\'opolis, SC, Brazil.}

\author{Luis C. N. Santos \orcidlink{0000-0002-6129-1820}}
\email{luis.santos@ufsc.br}

\affiliation{Departamento de F\'isica, CFM - Universidade Federal de Santa Catarina; \\ C.P. 476, CEP 88.040-900, Florian\'opolis, SC, Brazil.}

\author{D\'ebora P. Menezes \orcidlink{0000-0003-0730-6689}}
\email{debora.p.m@ufsc.br}

\affiliation{Departamento de F\'isica, CFM - Universidade Federal de Santa Catarina; \\ C.P. 476, CEP 88.040-900, Florian\'opolis, SC, Brazil.}

\author{Tobias Frederico \orcidlink{0000-0002-5497-5490}} 
\email{tobias@ita.br}

\affiliation{Departamento de F\'isica e Laborat\'orio de Computa\c c\~ao Cient\'ifica Avan\c cada e Modelamento (Lab-CCAM), Instituto Tecnol\'ogico de Aeron\'autica, DCTA, 12228-900, S\~ao Jos\'e dos Campos, SP, Brazil} 

\title{Strongly Interacting Quark Matter in Massive Quark Stars} 

\begin{abstract}
This paper investigates the properties of strongly coupled matter at high baryon densities (\(\rho_B\)) in quark stars (QSs). The QS model is based on the density-dependent quark mass (DDQM) framework, modified (MDDQM) by enhancing the single-gluon interaction to generate higher repulsive pressure. The model parameters are constrained using Bayesian inference, incorporating observational data from the pulsars HESS J1731$-$347, PSR J0030$+$0451, PSR J0740$+$6620, and PSR J0952$-$0607. Our results show that the MDDQM model produces QSs with higher mass and compactness compared to the DDQM model. Among the four MDDQM parameterizations studied, two yield maximum star masses of 1.86 and 2.10 \(\rm M_\odot\) and exhibit near-conformal behavior in the underlying quark matter (QM). The other two parameterizations, yielding QS masses of 2.30 and 2.37 \(\rm M_\odot\), correspond to a stronger interaction in the underlying QM. These findings provide important insights into the equation of state of deconfined QM and its implications for the structure and stability of QSs.
\end{abstract}

\maketitle

\section{Introduction}
In recent years, our understanding of the behavior of neutron stars (NSs) has been significantly enhanced due to the intense research activity in the field, motivated by the new body of observations. 
In particular, the recent direct detection of gravitational waves from binary NS merger events~\cite{LIGOScientific:2017vwq} and the data from the Neutron Star Interior Composition Explorer (NICER)~\cite{Gendreau2016} allow us to estimate the masses and radii of the NS simultaneously~\cite{riley2021nicer, riley2019nicer} with reasonable certainty. The observational evidence of heavier NSs of masses around 2$\rm M_\odot$ challenges our current understanding of how the nuclear equations of state (EoSs) can be stiffened adequately to support such masses against gravitational collapse. At the same time, the increasing information from QCD about how nuclear matter can dissolve into deconfined free quarks at higher baryon densities is gaining ground and improving our understanding of EoS under extreme conditions in NS interior~\cite{Annala:2023cwx, Annala:2019puf, Issifu:2023ovi, Baym:2017whm}.

These days, the characteristics of NS matter at the regions of its crust~\cite{Fortin:2016hny} have been fairly understood, thanks to extensive research on the topic~\cite{haensel2007neutron,chamel2008physics,bertulani2013neutron} spanning various research fields. For example, up to densities of about $\rho_{\rm CET} = 1.1\rho_0$, where $\rho_0=0.152 \rm fm^{-3}$ is the saturation density, using chiral effective field theory (CEFT), EoS was obtained with good accuracy~\cite{Gandolfi:2009fj, Tews:2012fj}. In this region, it has been established that matter exists in the hadronic phase. On the other hand, at higher $\rho_B$, perturbative-QCD (pQCD) modeled through high-energy phenomenology, treating matter with active quark and gluon degrees of freedom~\cite{Ivanenko1965, Ivanenko1969} turns out to give better results leading to EoSs with comparative accuracy at densities $\rho_{pQCD}\equiv \rho_B\gtrsim 40\rho_0$~\cite{Kurkela:2009gj, Gorda:2018gpy}. In these two limits, matter shows distinctly different properties~\cite{Lynn:2015jua, Drischler:2017wtt, Drischler:2020hwi, Keller:2022crb, Gorda:2021kme, Gorda:2021znl}. While high-density QM is nearly scale-invariant, hadronic matter, on the other hand, violates scale invariance due to chiral symmetry breaking~\cite{Fujimoto:2022ohj}.

Neutron stars allow us to explore the behavior of matter at extreme conditions (density and temperature) that cannot be created in conventional laboratories. At high baryon densities ($\rho_B\gg\rho_0$) in the NS core, for instance, the matter becomes highly compressed, and the quarks begin to overlap, causing the baryons to lose their identity and dissociate into QM. Such a scenario can result in two types of compact objects: a hybrid NS (an NS with a quark core)~\cite{Contrera:2022tqh, Lopes:2021jpm, Issifu:2023ovi} or a pure QS~\cite{Weber:2004kj, Zhang:2024xod} as a result of the high compression. While direct observational evidence for QS formation remains elusive, the most plausible formation pathways include NS phase transitions, direct quark-novae \cite{ouyed2002quark}, accretion-induced transitions \cite{Wiktorowicz:2017swq}, and compact star mergers \cite{Bombaci:2000cv, Most:2018eaw, Bauswein:2015vxa}. Advancements in gravitational wave astronomy and multi-messenger astrophysics are expected to provide stronger constraints on their existence.  To the best of our knowledge, the formation of QSs is theoretically expected for very massive and dense NSs, but there is no exact mass threshold that universally defines when an NS will transition into a QS.

Ivanenko and Kurdgelaidze hypothesized the existence of strange quark stars (SQSs) in 1965~\cite{Ivanenko1965, Ivanenko1970}, followed by Witten's conjecture~\cite{Witten:1984rs}, who suggested that strange quark matter (SQM) is possibly more stable than nuclear matter. Consequently, NSs can exist as SQSs~\cite{Melrose:2006zg}. This has motivated several QCD-inspired effective models to investigate the strongly coupled QM and QSs. Among the several models used in the literature are the quark-meson coupling model~\cite{Tsushima:1997cu}, quark mass density dependent model~\cite{PhysRevD.43.627, benvenuto1995strange}, confined density dependent model~\cite{peng2008deconfinement, PhysRevLett.86.3492}, Nambu-Jona-Lasinio (NJL) model~\cite{Buballa:1998pr, Schertler:1999xn, Klevansky:1992qe, Hatsuda:1994pi}, chiral SU(3) quark mean field model~\cite{Ping_2001}, Polyakov quark-meson coupling model~\cite{Schaefer:2007pw, Stiele:2013pma}, DDQM model~\cite{Peng:1999gh, Xia:2014zaa, daSilva:2023okq} and Polyakov extended NJL (PNJL) model~\cite{Costa:2010zw, Sakai1991}. Several studies have been carried out for pure SQM in $\beta$-equilibrium~\cite{li2020strange, zhang2021unified}, proto-strange stars~\cite{gupta2003study, PhysRevD.100.103012, Prakash:1996xs, Issifu:2023qoo} and hybrid stars with quark cores~\cite{Husain:2020nbb, Issifu:2023ovi}.

\begin{table*}[t]
\centering
        \caption{A table of six selected dimensional quantities determining the behavior of strongly interacting matter in five model frameworks. A similar table was published by Annala et al.~\cite{Annala:2023cwx} under the CC BY 4.0 license. }
    \begin{ruledtabular}
    \begin{tabular}{l c c c c c}
    
         \textbf{Quantity} &  \textbf{CEFT}\cite{Hebeler:2013nza, Landry:2020vaw}  &  \textbf{DNM}~\cite{Reed:2019ezm, Providencia:2023rxc} & \textbf{pQCD}\cite{Kurkela:2009gj, Gorda:2021znl} & \textbf{CFT}\cite{Kurkela:2009gj} & \textbf{FOPT}~\cite{Annala:2023cwx, Annala:2019puf}  \\
         \hline
         $c_s^2$  &  $\ll 1$ & $[0.25,\, 0.6]$  & $\lesssim 1/3$ & 1/3 & 0 \\

        $\Delta$ &   $\approx 1/3$  & $[0.05,\, 0.25]$  &  $[0,\, 0.15]$& 0 & $1/3-P_{\rm PT}/\varepsilon$  \\
        
         $\Delta'$ &   $\approx 0$ & $[-0.4,\, -0.1]$  &  $[-0.15,\, 0]$& 0 & $1/3-\Delta$  \\

         $d_c$  &  $\approx 1/3$ &$[0.25,\, 0.4]$  &  $\lesssim 0.2$ & 0 &$\leq 1/(3\sqrt{2})$  \\ 

        $\gamma$ &  $\approx 2.5$& $[1.95,\, 3.0]$  &  $[1,\, 1.7]$ & 1 & 0\\
        
        $P/P_{\rm free}$ &  $\ll 1$& $[0.25,\, 0.35]$  &  $[0.5,\, 1]$ & -- & $P_{\rm PT}/P_{\rm free}$
       
    \end{tabular}
    \end{ruledtabular}
    
    \label{tab1a}
\end{table*}

The measurement of NSs with masses $\rm M \geq 2M_\odot$ imposes a robust constraint on the EoS~\cite{ozel2016masses,riley2021nicer,antoniadis2013massive,demorest2010two,fonseca2016nanograv} which is so far not fulfilled by most phenomenological quark model EoSs. Equally important is the gravitational wave observation in the GW170817 event~\cite{LIGOScientific:2017vwq, LIGOScientific:2018cki} from binary NS merger and PSR J0030$+$0451~\cite{riley2019nicer}, both pointing to 1.4$\rm M_\odot$ NSs with radius $\rm R_{1.4} \lesssim 13.5 km$~\cite{Raaijmakers:2019dks}. 
Taking these constraints into account, together with the causality condition, $c_s^2\leq 1$, (where $c_s$ is the speed of sound in the unit of constant speed of light, set to $c=1$) requires that the EoS of dense matter (quarks and hadrons phases and only hadrons) changes faster from softness at densities $(1-2)\rho_0$ to relative stiffness at higher densities to reach the maximum stellar mass, which results in higher $c_s^2$. Additionally, at sufficiently high baryon densities found in the interior of NS ($(5-10)\rho_0$ or higher), $c_s^2$ is expected to approach the conformal limit ($c_s^2 = 1/3$) from below, which can be achieved in ultrarelativistic fluids with quark and gluon degrees of freedom. Such stringent constraints on NS matter rule out most phenomenological quark models because their EoSs are too soft~\cite{ozel2006soft,rodrigues2010quark, Albino:2021zml} to satisfy the recent mass constraints. Particularly the measurement of stars with $M \sim 2.5 \rm M_\odot$~\cite{LIGOScientific:2020zkf,abbott2020gw190425} and $M > 2.27 \rm M_\odot$~\cite{linares2018peering,romani2022psr}, calls for further modifications of the existing models to achieve such masses.

We usually rely on the signatures the dense matter leaves on various thermodynamic properties through the EoS to distinguish between different states of matter. For instance, the speed of sound squared takes on a constant value of \(c_s^2 = 1/3\) in exactly scale-invariant matter. This value is expected to be approached slowly from below in high-density QM. The normalized trace anomaly $\Delta$~\cite{Fujimoto:2022ohj}, its logarithmic change $\Delta'$, the polytropic index $\gamma$, the matter pressure normalized by the pressure of the free non-interacting quarks $P/P_{\rm free}$ (where $P$ is pressure and $P_{\rm free}$ is the free massless, non-interacting Fermi Dirac pressure), the effective running coupling constant $\alpha_s$ and the $\beta$-function are among some of the quantities whose characteristics are used to distinguish matter states. The numerical values of these properties determined through five different model frameworks at different $\rho_B$ have been presented in Tab.~\ref{tab1a} for comparison.

Furthermore, at very high $\rho_B$ the QM is weakly coupled and shows approximately conformal invariant behavior {following the expectation of asymptotic freedom and restoration of chiral symmetry}. That notwithstanding, the formation of condensates through diquarks and paired quarks~\cite{Baym:2017whm}, subdominant loop effects, and a small fraction of $\rm u,\, d\; and \; s$ (up, down and strange quarks respectively) quark masses may subtly violate the conformality in this region. From Tab.~\ref{tab1a}, matter in {the high $\rho_B$ regime} will have a small positive value of $\Delta$, a small negative value of $\Delta'$, $c_s^2\lesssim 1/3$ and $1\leq \gamma\leq 1.7$. These are typical values obtained from ultrarelativistic systems significantly different from the ones obtained from the hadronic phase. In the hadronic phase, low $\rho_B$ {\it ab-initio} results have been precisely determined~\cite{Hebeler:2013nza, Tews:2012fj}. On the other hand, at higher $\rho_B$, phenomenological models are adopted in this case, the dominant nucleon mass scales strongly break the conformal symmetry~\cite{Annala:2019puf, Malik:2023mnx} yielding properties significantly different from the ones observed from ultrarelativistic systems.

{This work aims to extend the QM model, specifically the DDQM model~\cite{Peng:1999gh, Xia:2014zaa, daSilva:2023okq}, to reproduce more compact QSs} and investigate the properties presented on Tab.~\ref{tab1a}.  
Using QSs~\cite{Ivanenko1965, Ivanenko1970} as the medium for this investigation, we are interested in the size of the star and how QM comports itself inside it.
According to~\cite{ozel2016masses}, we know the mass of $\sim 35$ NSs with good precision, the values of these masses are between $1.17$ and $2.0$M$_{\odot}$, and we know the radius of $\sim 12$ NSs, in the range $10-11.5$km. Two of the more important mass determinations are $1.928^{+0.017}_{-0.017}$M$_{\odot}$ for PSR J1614$-$2230~\cite{demorest2010two,fonseca2016nanograv} and $2.01^{+0.04}_{-0.04}$M$_{\odot}$ for PSR J0348$+$0432~\cite{antoniadis2013massive}, which were the first measurements with good precision to confirm that a NS could reach $2$M$_{\odot}$. Besides, in~\cite{Weber:2004kj} we can find a list of known compact stars that present some properties that make them promising candidates to be SQS. In determining the free parameters of the model, {the recent mass and radius data of observed pulsars}: PSR J0952$-$0607~\cite{romani2022psr}, PSR J0740$+$6620~\cite{riley2021nicer}, PSR J0030$+$0451~\cite{riley2019nicer}, and HESS J1731$-$347~\cite{doroshenko2022strangely} were used, in addition to the Bodmer and Witten conjecture~\cite{Witten:1984rs, Bodmer:1971we, Glendenning2012}, as the benchmark for determining stable QSs. The conjecture states that the energy density, $\varepsilon$, per $\rho_B$ of a SQM at the surface of the star ($P=0$) should be less than the energy per nucleon, $E/A$, of $^{56}$Fe, \textit{i.e.}, $(\varepsilon/\rho_B)_{\rm SQM}\leq 930$MeV. Simultaneously, a two-flavor quark system (up and down QM system) must satisfy $(\varepsilon/\rho_B)_{\rm 2QM} > 930$MeV. Otherwise, protons and neutrons would dissociate into their constituents u and d quarks. {When both conditions are satisfied we say that the star is inside the stability window.}

The paper is organized such that, in Sec.~\ref{m} we present the model intended for the study and discuss its important properties in six subsections. In Subsec.~\ref{md} we discuss the proposed modification to the model and proceed to do the thermodynamic consistency analysis in Subsec.~\ref{t}. In Subsec.~\ref{pdm}, we discuss the properties of the QM suitable for application to QSs, the behavior of particle degrees of freedom with $\rho_B$ was presented in Subsec.~\ref{qd}. The conformal properties of the QM that enable us to differentiate between confined and deconfined QM states were presented in Subsec.~\ref{cp}, and applications of the model to QSs were presented in Subsec.~\ref{acs}. In Sec.~\ref{ra}, we present our findings and analyze them in detail and our final remarks are in Sec.~\ref{fr}.

\section{The Model}\label{m}

The interior of NSs, assumed here to consist of SQM, approaches densities of $\rho_B \sim (2-10)\rho_0,$ beyond the reach of \textit{ab initio} QCD calculations. While quarks and gluons become relevant at these densities, perturbative QCD (pQCD) is not yet applicable, necessitating phenomenological models like the density-dependent quark mass (DDQM) model~\cite{Peng:1999gh, Wen:2005uf, Chen:2021fdj} to describe quark matter (QM).  

The DDQM model follows from Hamiltonian for the effective quark degrees of freedom in the matter given by
\begin{equation}
    H_{\rm Q} = H_k + \sum_{i = u,\,d,\,s}m_{i0}\bar{q}q + H_I,
\end{equation}
and its equivalent,
\begin{equation}\label{heqv}
    H_{\rm eqv} = H_k + \sum_{i = u,\,d,\,s}m_{i}\bar{q}q,
\end{equation} 
where $H_k$ is the kinetic term, $H_I$ is the interacting term, $\bar{q}$ and $q$ are the quark fields, $m_{i0}$ is the current quark mass, and $m_i$ is the equivalent quark mass.  {Having that $ H_{\rm Q}=  H_{\rm eqv}$, the equivalent quark mass in Eq.~(\ref{heqv}) is given by:
\begin{equation}
    m_i(\rho_B) = m_{i0} + m_I(\rho_B),
\end{equation}
with $m_I(\rho_B)$ representing the interacting part of $m_i(\rho_B)$ parameterizing the effect of the baryonic density.} The originally proposed ansatz for the cubic mass scaling formula of the $m_i(\rho_B)$ given in~\cite{Peng:1999gh} reads: 
\begin{equation}\label{d}
    m_i = m_{i0} + \dfrac{D}{\rho_B^{1/3}}  = m_{i0} + m_I(\rho_B).
\end{equation}
Confinement is modeled as quarks acquiring infinite mass in a vacuum, analogous to the MIT bag model~\cite{thomas1984chiral}. At asymptotic densities, $\rho_B \rightarrow \infty$, the $m_i$ approaches $m_{i0}$. 

For astrophysical applications, quark masses must support \(\geq 2M_\odot\) stars~\cite{riley2021nicer, romani2022psr}. However, Eq.~\eqref{d} introduces an attractive pressure \(P_i \sim \rho_B^2 d m_i / d \rho_B\), limiting the maximum mass. To counteract this, an additional repulsive term is included~\cite{Xia:2014zaa}:  
\begin{equation}\label{d1}
    m_i = m_{0i} + \frac{D}{\rho_B^{1/3}} + C\rho_B^{1/3},
\end{equation}  
where \(C\) represents one-gluon exchange strength. This modification allows QM to satisfy NS mass constraints while maintaining quark confinement at low densities.  

Eq.~\eqref{d1} mirrors the Cornell potential for heavy-quark confinement~\cite{eichten1978charmonium}:  
\begin{equation}
    V(r) = -\frac{\beta}{r} + \sigma r + V_0,
\end{equation}  
where \(\beta\) and \(\sigma\) represent deconfinement and confinement strengths, respectively. Using \(r \sim 1/\rho_B^{1/3}\), Eq.~\eqref{d1} naturally emerges. The sign of $C$ determines the model’s physical behavior:  
$C < 0$ (attractive) favors lower-mass quark stars~\cite{Chen:2021fdj}.  
$C > 0 $ (repulsive) stabilizes heavier stars~\cite{rodrigues2010quark, Albino:2021zml, alford2007quark, franzon2012self, klahn2015vector, Lopes_2021, song2019effective, otto2020nonperturbative}. For NSs satisfying the \(2M_\odot\) constraint, repulsive interactions prevent collapse, making the DDQM model a viable approach for astrophysical QM studies.

\subsection{The Modified DDQM model}\label{md}
The main motivation for this modification is to achieve an enhanced maximum stellar mass with smaller radii and, consequently, a more compact QS than the one that can be achieved from Eq.~(\ref{d1}). In~\cite{daSilva:2023okq}, the authors compared QSs built from Eq.~(\ref{d1}) to others built from the vector MIT bag model~\cite{PhysRevD.9.3471, Serot1992}, {which builds upon the original MIT bag model~\cite{MITbag} by integrating certain aspects of the quantum Hadrodynamics (QHD)~\cite{Cierniak:2018aet, Lopes_2021, Lopes_2021b},} and they observed that the DDQM model produces less compact QSs with masses far lower than that of the vector MIT bag model. Using Bayesian inference to fix the free model parameters in an optimized manner, imposing the recent astrophysical constraints~\cite{daSilva:2023okq}, the maximum mass reached for the DDQM model was $\sim 2.18\rm M_\odot$. On the other hand, the vector MIT bag model reached $\sim 2.54\rm M_\odot$ for the same constraints. Other studies of QSs using the DDQM model yield maximum masses less than the ones obtained from the vector MIT model, particularly when the quark masses reported in the Particle Data Group (PDG) are used~\cite{Chen:2023rza, Marzola:2023tdk, Xia:2014zaa}. In~\cite{Backes:2020fyw}, the authors obtained a slightly higher maximum mass up to $\sim 2.37\rm M_\odot$ using arbitrary current quark masses. Therefore, the model is sensitive to the current quark masses within NS densities. We adopted the data for current quark masses reported in the PDG~\cite{ParticleDataGroup:2022pth} for this work, similar to~\cite{daSilva:2023okq}.

From Eq.~(\ref{d1}), we know that the term {$D\rho_B^{-1/3} $ contributes to a negative} pressure that acts to reduce the maximum stellar mass and the {$C\rho_B^{1/3}$ term contributes to a positive pressure}  that acts to augment the maximum stellar mass. So, we introduce an extra density dependence term to $C\rho_B^{1/3}$ by including $\kappa\rho_B^{2/3}$ and set $C=1$, yielding
\begin{equation}\label{a1}
    m_i = m_{i0} + \dfrac{D}{\rho_B^{1/3}} + \Big(1+\kappa\rho_B^{1/3}\Big)\rho^{1/3}_B,
\end{equation}
where $D$ and $\kappa$ are constants with dimensions [MeV$^2$] and [MeV$^{-1}$] respectively. During our initial tests with MDDQM, we kept $C$ as a parameter in our model and conducted several tests. However, we consistently observed that $C$ approaches unity across all the tests. Therefore, to simplify the model and reduce the number of free parameters, we subsequently fixed $C$ at 1. The primary motivation for modifying the vector interaction in the DDQM model is to enhance the repulsive pressure within the star, enabling it to support a larger QS mass. This modification is crucial for stabilizing the star against gravitational collapse, leading to higher maximum masses and potentially influencing its compactness, depending on the balance between repulsion and gravitational pull. 

A similar idea was used in modeling the vector MIT bag model~\cite{PhysRevD.9.3471, Serot1992} where the authors aimed at producing stars with higher maximum masses. In relation to the DDQM compared to the NJL (Nambu–Jona–Lasinio) model, we can naively compare the original DDQM expression in Eq.~(\ref{d})  with the NJL model. Here, we retain \( m_{0i} \) at asymptotic densities, which mimics the NJL model at high densities. In the NJL model, the dynamically generated mass due to spontaneous chiral symmetry breaking evolves with density, while the current quark mass remains unchanged \cite{RevModPhys.64.649}. At sufficiently high densities, chiral symmetry is restored when the dynamically generated mass vanishes, leaving only the current mass. However, introducing vector and mean-field interactions in the NJL model significantly alters the behavior of dense quark matter. These effects 
stiffen the EoS, and allow for more massive quark stars \cite{Klahn:2015mfa, Buballa:2003qv}. In this context, the DDQM in Eq.~(\ref{d1}) and the MDDQM in Eq.~(\ref{a1}) can be interpreted analogously: the additional terms \( C\rho_B^{1/3} \) and \( \kappa\rho_B^{2/3} \) represent a collective contribution at higher densities mimicking the vector and mean-field effects in the NJL model, even though \( C\rho_B^{1/3} \) and \( \kappa\rho_B^{2/3} \) were inspired by the one-gluon exchange interaction.

In Eq.~(\ref{a1}), the value and the sign of $\kappa$ are sensitive to the properties of the stellar matter. For instance, a negative $\kappa$ will reduce the pressure and, consequently, the compactness of the star and vice versa. The nonrelativistic static potential between two heavy quarks ($\bar{Q}Q$) with masses $m_Q\gg \Lambda$, where  $\Lambda$ is the QCD scale, can be expressed as
\begin{equation}\label{a2}
    V(r) = \kappa_s\dfrac{\alpha_s(r)}{r}+\sigma r + V_0, 
\end{equation}
with constant coupling, $\alpha_s$~\cite{Deur:2016tte} and a color factor $\kappa_s$ (it can be positive or negative)~\cite{Liu:2019zoy, Debastiani:2017msn}.  In relation to the logarithmic QCD running coupling 
\begin{equation}
    \alpha_s(Q^2) = \dfrac{4\pi}{\beta_0 \ln{(Q^2/\Lambda^2)}},
\end{equation}
where $Q$ is a space like momentum, $\Lambda$ is the QCD parameter and $\beta_0 = 11-\frac{2}{3}N_f$ is the one-loop beta function coefficient for $N_f$ flavors. By defining the momentum as $Q\sim 1/r$ we have,
\begin{equation}
    \alpha_s(r) \approx \dfrac{4\pi}{\beta_0 \ln(1/(\Lambda r)^2)}.
\end{equation} 
In a simple quark model like the one presented in Eq.~(\ref{a2}), the first term is the perturbative Coulomb-type single-gluon exchange contribution, which is sensitive to the hadron wave function, the fine structure, and the hadron spectrum~\cite{Eichten:2002qv} with $\kappa_s = -(4/3)$. On the other hand, considering quark-antiquark pairs or multi-quark systems~\cite{Huang:2023jec, Martens:2006ac}, in some color configurations, the interaction can be repulsive~\cite{Wang:2023igz}. For instance, in the color octet state, the interaction can be expressed as
\begin{equation}
    V_{\rm octet}(r)=\dfrac{1}{6}\dfrac{\alpha_s(r)}{r},
\end{equation}
which is repulsive with $\kappa_s = 1/6$. These interactions are indeed expected to dominate from 4-quark systems ($qq\bar{q}\bar{q}$) to a regime where there are large numbers of quark and anti-quark pairs leading to the formation of clusters. Moreover, at high \(\rho_B\), quark-quark pairs can exist in two possible color configurations: The color anti-triplet (\(\bar{3}\)) and the color sextet (6). In the context of this discussion, where a repulsive interaction is required to support heavy QSs, the sextet configuration is ruled out because it corresponds to an interaction strength of \(\kappa_s = -1/3\), leading to an attractive potential. Conversely, the anti-triplet (\(\bar{3}\)) configuration is favored, as it has an interaction strength of \(\kappa_s = 2/3\), which results in a repulsive potential. This repulsion can stabilize and facilitate the formation of diquark states, which are relevant in the context of QSs. For the sake of our analysis, we will use $\kappa_s = 1/6$ through out the paper.

Comparing Eqs.~(\ref{a1}) and~(\ref{a2}), we can infer that a strong vector repulsion is required between the quarks in the dense quark medium to justify how the quark core can support the $2$M$_\odot$ threshold for the NSs. Additionally, in Eq.~(\ref{d1}), $C$ was determined to have positive values, as shown in Tab.~\ref{tabdd}, to fit the current observational data (see Refs.~\cite{daSilva:2023okq, Backes:2020fyw} and references therein).  Therefore, repulsive interaction between quark pairs in high-density QM was long envisaged in the study of QSs. This leads to the identification of $\alpha_s(r)$ with $1+\kappa \rho_B^{1/3}$ by comparing Eqs.~(\ref{a1}) and (\ref{a2}), which we refer to as the `effective running coupling' here and hereafter. Thus, we have:
\begin{equation}\label{alphas}
    \alpha_s(\rho_B) = \kappa_s\Big(1 + \kappa \rho_B^{1/3} \Big),
\end{equation}
when we use the ansatz, $r\sim 1/\rho_B^{1/3}$. This implies that in the DDQM, $C \approx \alpha_s(0)$, which is a common assumption in Cornell potential models \cite{Eichten:2002qv, PhysRevLett.34.369}. In the MDDQM framework, we defined $C$ as a function of $\rho_B$, i.e., $C(\rho_B)= 1+\kappa \rho_B^{1/3}$ leading to the variation of $\alpha_s$ with $\rho_B$. Through this expression, we can calculate the so-called $\beta$-function as a function of $\rho_B$ using the relation
\begin{equation}\label{beta}
    \beta(\alpha_s(\rho_B) ) := \rho_B^{1/3} \dfrac{d\alpha_s(\rho_B) }{d\rho_B^{1/3}} := \dfrac{d\alpha_s(\rho_B) }{d\ln \rho_B^{1/3}},
\end{equation}
where $Q\sim \rho_B^{1/3}$, with $Q$ the spatial momentum. Thus, the sign of $\kappa$ significantly influences the behavior of $\alpha_s$ and $\beta$. For instance, the $\beta$-function is known to be negative, and $\alpha_s$ is positive but decreases with $\rho_B$ in pQCD, that behavior can only be achieved {in the expressions of Eqs.~(\ref{alphas}) and~(\ref{beta})} if $\kappa<0$. That notwithstanding, further analyses (it will be clear in subsequent sections) show that $\kappa>0$ is the appropriate choice for constructing massive QSs within the $2$M$_\odot$ threshold determined through NS observation data. The analysis above suggests that as the density increases, many gluons are exchanged between the densely packed quarks, even in the weak coupling regime, in a way that produces a repulsive behavior from the colored gluonic field. The DDQM model is just an effective means of including such physics. The densely packed quarks can hardly recoil, and thus the gluons in flight should carry small momentum fractions, however, the gluon densities presumably become so large that they saturate, leading to the dominance of the classical nonlinear dynamics. This phenomenon follows the analogy of gluon saturation phenomena, namely the ``color glass condensate" (CGC), which is crucial for understanding the initial conditions in high-energy collisions~\cite{McLerran:1993ni, McLerran:1993ka, Iancu:2000hn}.

\subsection{Thermodynamic Consistency}\label{t}
The DDQM model is used to study SQM, where $\rho_B$ serves as the medium for interactions between valence quarks. 
Quark mass models depending on chemical potential \(\mu_i\), \(\rho_B\), and temperature are discussed in~\cite{Zhang:2021qhl, Goloviznin1993, Backes:2020fyw, Wen:2005uf, Chu:2021aaz, Schertler:1996tq, Peshier:1999ww}. Initial concerns about quark mass scaling and thermodynamic consistency have been addressed in~\cite{Backes:2020fyw, Chen:2021fdj, Peng:2000ff, Peng:1999gh}, allowing the model’s application.

To ensure thermodynamic consistency, the free quark mass \(m_{0i}\) in the Helmholtz free energy formula is replaced by \(m_i(\rho_B)\), which depends on density. The real chemical potential \(\mu_i\) is replaced by an effective chemical potential \(\mu_i^*\).  After establishing consistency, other thermodynamic quantities, like pressure and energy density, are derived using \(\mu_i^*\), a full discussion is in \autoref{ap1}.

\subsection{Properties of the Dense Stellar Matter}\label{pdm}
The QM is composed of three flavor quarks $\rm u,\, d,\; and\; s$ in $\beta$-equilibrium with electrons $e$. Following the equilibrium reactions; $d \rightarrow u + e^- + \bar{\nu}_e$, $u + e^- \rightarrow d + \nu_e$, $s \rightarrow u + e^- + \bar{\nu}_e$, and $u + e^- \rightarrow s + \nu_e$, we derive a relation between the effective chemical potentials and the electron chemical potential $\mu_e$
\begin{equation}
    \mu_u^* + \mu_e = \mu_d^* = \mu_s^*,
\end{equation}
where the subscripts represent the individual particles present in the system.  It is important to mention that this $\beta$-equilibrium condition can also be expressed in terms of $\mu_i$ for the individual particles as discussed in~\cite{Backes:2020fyw}. {However, one should keep track of the $\mu_I$ that appears in the $\mu_i^*$ as a result of the thermodynamic consistency of the model, that connects it to $\mu_i$ as discussed in the previous subsection}. Moreover, we imposed charge neutrality conditions on the stellar matter through 
\begin{equation}
    \frac{2}{3} \rho_u - \frac{1}{3} \rho_d - \frac{1}{3} \rho_s - \rho_e = 0.
\end{equation}
Since the quark mass formula in Eq.~(\ref{d}) is flavor-independent, the degeneracy factor is $g_i=6$ (3 colors $\times$ 2 spins). 

\subsection{Quark Degrees of Freedom at Higher $\rho_B$}\label{cdst}\label{qd}

At high \(\rho_B\), quark degrees of freedom emerge, as nucleons interacting via a static potential can no longer describe the system. As \(\rho_B\) increases, QM undergoes percolation, allowing quarks to propagate freely~\cite{Baym:1979etb, Celik:1980td, Satz:1998kg}, similar to how electrons become itinerant in a compressed atomic gas. At low densities (\(\rho_B \lesssim 2\rho_0\)), hadronic matter consists of nucleons and pions, with interactions mediated by mesons or quarks~\cite{Drischler:2021kxf}. At intermediate densities (\(2\rho_0 \lesssim \rho_B \lesssim (4-7)\rho_0\)), colored quarks and diquarks appear, leading to deconfined quark matter~\cite{Baym:2017whm}. At higher densities (\(\rho_B > (4-7)\rho_0\)), QM becomes fully percolated, with quarks and diquarks forming local color-neutral structures. pQCD applies at extreme densities (\(\rho_B \gtrsim 40\rho_0\))~\cite{Annala:2023cwx, Kurkela:2009gj}. As \(\rho_B\) increases, the system undergoes phase transitions, and QCD condensates lower the energy, breaking QCD symmetries~\cite{Pethick:2015jma}, significantly affecting the energy density in the hadronic matter and NS interiors.

Chiral symmetry breaking, where quark condensates (\(\langle \bar{q}q \rangle\)) give rise to Nambu-Goldstone bosons like pions and kaons~\cite{PhysRev.122.345}, leads to color superconductivity at high \(\rho_B\)~\cite{Son:1999cm, Fukushima:2004bj, Yamamoto:2007ah, Baym:2017whm}. 
At high $\rho_B$, QM is theoretically expected to undergo key phenomena, including deconfinement, distinct QM phases, color superconductivity, chiral symmetry breaking, and pQCD effects. Among these, deconfinement, QM phases, and the transition from hadronic to QM have the highest likelihood of being observed in QSs. Indirect evidence of color superconductivity may also emerge from cooling rates and transport properties. Chiral symmetry restoration is expected, though its direct observability in QSs is uncertain and likely inferred from thermodynamic behavior. In contrast, pQCD effects are more relevant to high-energy heavy-ion collisions and are unlikely to manifest in QSs under typical conditions. Interested readers may refer to Ref.~\cite{Baym:2017whm} for a comprehensive review of the behavior of QM under extreme density conditions and its implications for NSs and QSs.

\subsection{Conformal Properties of the QM}\label{cp}
In QCD at low energy, conformal transformations lead to the persistence of the dilatation current $s_D^\mu$, whose divergence gives the trace anomaly $\Theta$, i.e., $\partial_\mu s_D^\mu = T^\mu_\mu= \Theta$. In classical gluodynamics, where the theory is conformally invariant, $\Theta=0$; however, in QCD, both the quark masses, gluon condensate, and trace anomaly break this symmetry;
\begin{equation}
    \Theta = \dfrac{\beta}{2g}F_{\mu\nu}^aF^{\mu\nu}_a + (1+\gamma_m)\sum_fm_f\bar{q}_fq_f
\end{equation}
where $\beta/(2g) = -(11-2N_f)(\alpha_s/(8\pi)) + {\cal O}(\alpha_s^2)$ is the QCD $\beta$-function, with $\alpha_s$ the strong coupling constant and the flavor number $N_f$, the anomalous dimension of the quark mass $\gamma_m = 2\alpha_s/\pi+ {\cal O}(\alpha_s^2)$ and the strong coupling constant $g$. At a finite temperature ($T$) and/or baryon chemical potential ($\mu_B$), the expectation value of $\Theta$ has both matter and vacuum contributions $\langle \Theta\rangle = \langle \Theta\rangle_{T,\mu_B} + \langle \Theta\rangle_0$, with $\langle \Theta\rangle_0$ being the vacuum expectation value at $T=\mu_B=0$. In the current work, we focus on the matter part where 
\begin{equation}
    \langle \Theta\rangle_{\mu_B} = \varepsilon - 3P,
\end{equation}
with the energy density $\varepsilon$ and pressure $P$. This follows directly from the energy-momentum tensor trace; by convention $\langle \Theta\rangle_{\mu_B}$ is also referred to as the trace anomaly. It can be shown through thermodynamic properties that, at higher densities where quarks and gluons are expected to be in a deconfined state, $\langle \Theta\rangle_{\mu_B}\rightarrow 0$, $P \propto \mu_B^4$ corresponding to $\varepsilon \approx 3P$, for a strongly coupled conformal matter. At that density, the conformal symmetry of the theory is expected to be restored, approximately. 

The proposed measure of trace anomaly in NSs, as shown in~\cite{Fujimoto:2022ohj, Marczenko:2022jhl}, is determined by scaling $\langle \Theta\rangle_{\mu_B}$ with  
$\varepsilon$ in the form 
\begin{equation}\label{dp}
    \Delta \equiv \dfrac{\langle \Theta\rangle_{\mu_B}}{3\varepsilon} = \dfrac{1}{3}- \dfrac{P}{\varepsilon}.
\end{equation}
Ensuring thermodynamic stability $P>0$ and causality $c_s^2\leq 1$, with $c_s$ the speed of sound in the unit of constant speed of light $c$, $\Delta$ lies in the range $-2/3\leq \Delta < 1/3$. The $\Delta$ is the normalized form of the QCD trace anomaly. It measures the degree of conformal symmetry in a superdense matter. The conformality is expected to be fully satisfied when $\Delta =0$ at extremely high densities following pQCD predictions~\cite{Fujimoto:2022ohj}. Its behavior at intermediate densities, reachable in the NS interior, is still under intense research. However, the information extracted from $\Delta$ through analyzing NS observable data is still EoS model-dependent. It has been used to study the possible conformality of NS matter in~\cite{Marczenko:2022jhl}, and the presence of quark cores in hybrid NSs in~\cite{Annala:2023cwx}. We intend to use this parameter to study near-conformality in dense QM comparing it with other parameters like the $c_s$ whose properties are relatively well known to make inferences. One of the model-independent ways for determining $\Delta$ in NS matter has been proposed in~\cite{Cai:2024oom}, using the central energy density ($\varepsilon_c$) and its corresponding pressure $P_c$ of the observed stars. As a result, the central trace anomaly $\Delta_c$ can be measured using $\Delta_c = 1/3 - P_c/\varepsilon_c$ in a model-independent manner using NS observed data.

From this relation, we can define the logarithmic rate of change connecting $\varepsilon$ and $\Delta$ as $\Delta' \equiv d\Delta/d\ln \varepsilon$ and the polytropic index $\gamma$ which is also a measure of conformality of a strong interacting matter given as $\gamma = d \ln P/d\ln \varepsilon$. The $c_s$ expressed in terms of $P$ and $\varepsilon$ as $c_s^2 = dP/d\varepsilon$ can then be written in terms of $\Delta$ as 
\begin{equation}
    c_s^2 = \dfrac{1}{3} - \Delta - \varepsilon \dfrac{d\Delta}{d\varepsilon}.
\end{equation}
The scale invariance of the theory is restored at $\Delta \rightarrow 0$, which corresponds to $c_s^2 = 1/3$, this is well known in the literature as the conformal invariant limit~\cite{Kurkela:2009gj, Altiparmak:2022bke, Annala:2019puf, Annala:2023cwx}. The $\Delta'$ can be expressed in terms of $c_s^2$ as
\begin{equation}\label{dp1}
    \Delta' = \dfrac{P}{\varepsilon}-c_s^2,
\end{equation}
therefore, $\Delta'$ lies within the range $\rm -1/3<\Delta'\leq 2/3$. The $\Delta$, $\Delta'$, $\gamma$, and $c_s^2$ have different numerical values at low and higher $\rho_B$, and in approximately conformal QM at asymptotically high $\rho_B$.  Vanishing $\Delta$ occurs in conformal limit, thereby serving as a measure of a property of the strongly coupled QM aside from the $c_s^2$ and the $\gamma$ as discussed in~\cite{Annala:2019puf, Annala:2023cwx}. To be able to classify between non-conformal and approximately conformal matter, we further calculate a new quantity $d_c$ that combines $\Delta$ and $\Delta'$ in a single expression
\begin{equation}
    d_c = \sqrt{\Delta^2 +( \Delta')^2}.
\end{equation}

 We present the theoretical limits for $\Delta$, $\Delta'$, $\gamma$, $c_s^2$ and $d_c$ on Tab.~\ref{tab1a} for CEFT, DNM (dense nuclear matter), pQCD, Conformal Field Theory (CFT) and First Order Phase Transition (FOPT) for comparison. {The value of $d_c$ at lower densities, when we consider that the pressure on the surface of the star is $P=0$, gives $\Delta=1/3$ and $\Delta' = 0$, thus at low densities $d_c$  $\approx 1/3$, as can be seen in Tab.~\ref{tab1a} for the CEFT column. For the core of the star, we resort to the values for DNM that estimate $0.25 \leq d_c \leq 0.4$ in the column, DNM of Tab.~\ref{tab1a}. As for the possibility of a phase transition, we know from FOPT that $c_s^2=\gamma=0$, which leads to $d_c \leq 1/ (3 \sqrt{2}) \approx 0.2357 $.} This was adopted in~\cite{Annala:2023cwx} as their criterion, setting $d_c < 0.2$ for identifying near-conformal matter at NS densities so that FOPTs are not confused with conformalized matter.
These quantities are related to each other through this set of equations 
\begin{align}
  \Delta' = c_s^2\Big( \dfrac{1}{\gamma} - 1\Big).
\end{align}
It is easier to analyze the effect of one quantity on the other through this set of equations. Additionally, the free massless, non-interacting Fermi-Dirac pressure, given by 
\begin{equation}
    P_{\rm free}(\mu_B)= \dfrac{3}{4\pi^2}\Bigg(\dfrac{\mu_B}{3}\Bigg)^4,
\end{equation}
valid for a system of three quarks is used to normalize the $P$. This parameter does not necessarily determine the conformality of matter, but it determines the effective degrees of freedom of weakly coupled and strongly coupled matter~\cite{Cardy1996, Gubser:1998nz}. The theoretical estimates of $P/P_{\rm free}$ for various models are also recorded in Tab.~\ref{tab1a}.  The equations derived here are applied to the EoSs of the MDDQM model, and the results are plotted in Fig.~\ref{fig9} for comparison with the standard values presented in Tab.~\ref{tab1a} and also marked on the graph.

The parameters discussed provide critical insights into the EoS and the properties of QM in compact stars. The \( c_s^2 \) serves as a model-independent probe of EoS stiffness, phase transitions, and interaction strength, with values exceeding \( 1/3 \) indicating strong interactions. The \( \Delta \) quantifies deviations from conformality, distinguishing weakly from strongly interacting QM and influencing the structure of QSs and NSs. The logarithmic rate of change, \( \Delta' \), captures the evolution of the \( \Delta \) with \(\varepsilon\), signaling instabilities, phase transitions, or deconfinement. The deviation from conformality parameter, \( d_c \), highlights the extent of non-conformality, with larger values suggesting stronger interactions. The \( \gamma \) measures the relationship between pressure and energy density, offering a model-independent indicator of conformality. Finally, the pressure ratio \( P/P_{\rm free} \) assesses the deviation from an ideal quark gas, with values far from unity suggesting strong interactions, phase transitions, or quark confinement, while values near one indicate weakly interacting or deconfined QM.

These metrics are crucial for understanding the nature of QM in compact objects and their thermodynamic behavior. In QSs and NSs, multiple properties must be analyzed simultaneously to infer accurately whether QM is near-conformal or strongly interacting. Specifically, \(\Delta\), \(\Delta'\), and \(d_c\) reflect the degree of conformality, with stiffer EoS often associated with near-conformal matter. However, stiffness alone does not guarantee conformality. Meanwhile, \(c_s^2\) and \(\gamma\) capture the local behavior of the EoS, while \(P/P_{\rm free}\) quantifies how the pressure deviates from an idealized free quark gas, providing insight into interaction effects and degrees of freedom in QM. 

\subsection{Application to Compact Stars}\label{acs}
QSs are compact objects composed of QM consisting of up, down, and strange quarks. Recent observational advancements have provided significant insight into the nature of dense matter in the stellar interior. In this sense, QSs are expected to have a distinct mass-radius relation compared to NS, and it is expected that the study of strongly coupled matter at high densities can reveal properties between the mass and radius of a QS that can be compared with recent observations. {One of the highlighted differences between the NSs and QSs is that QSs may be either bare or contain a crust composed of ionized atoms~\cite{Kettner1995, melrose2006pair}. Despite the counterarguments regarding the existence of a crust in SQS~\cite{Kettner1995,usov1997low,melrose2006pair}, the crust could be blown away during the formation stages of the star~\cite{Melrose:2006zg}, so their structure is generally constructed without the crust.}  Assuming {static, spherically symmetric cold stars}, we can use the Tolman--Oppenheimer--Volkoff (TOV) equation to describe a QS \cite{PhysRev.55.374}, see \autoref{appendix:TOV} for the TOV equations.


In the context of observations of compact stars, in addition to the star's mass, another important parameter in the measurement is the so-called tidal deformability associated with the deformation of the shape of each star in a binary system. In this case, due to the external field $\epsilon_{ij}$ of its companion, there is a quadrupole moment $Q_{ij}$ of the form 
\begin{equation}
 Q_{ij} =  -\lambda\epsilon_{ij},
 \label{eqq3}
\end{equation}
where $\lambda$ is 
the tidal deformability parameter, also called the tidal Love number. The full expression of the dimensionless tidal deformability and how it is applied in the case of QSs is contained in \autoref{ap}.

We are interested in analyzing the constraints imposed by recent observations. In this regard, millisecond pulsars such as PSR J0740$+$6620 and PSR J0952$-$0607 can be used to constrain models to calculate the EoS associated with QSs. Due to their higher maximum masses of around $\rm 2M_\odot$, these systems are difficult to describe with some QM models. On the other hand, the low-mass compact stars HESS J1731$-$347 and PSR J0030$+$0451, require a particular configuration of the EoS parameters. Therefore, considering a modified DDQM model, we use Bayesian inference to find the best set of parameters that satisfy the constraints from all the compact stars studied.

\section{Results and Analysis}\label{ra}


\begin{table*}[t]
 \centering
        \caption{QS properties for the different parameter sets of the DDQM model and MDDQM models analyzed. In this Table, $\rho_c$ is the central baryon density determined at the center ($r=0$) of a star with a maximum mass $\rm M_{max}$ in each parameterization. Moreover, $R_{1.4}$ is the radius of the canonical star of mass 1.4 $\rm M_\odot$, $\Lambda_{1.4}$ is its dimensionless tidal deformability parameter, and $\varepsilon_c$ is the central energy density.}
    
    \begin{ruledtabular}
    
    \begin{tabular}{c c c c c c c c c c|}
    \multicolumn{7}{c}{DDQM model Parameters~\cite{daSilva:2023okq}(previous work)} \\
    \hline
         Scenario & $C$ & $\sqrt{D}[{\rm MeV}]$ & M$_{\rm max}[{\rm M}_{\odot}]$ & $R[{\rm km}]$ & $\rho_c[{\rm fm}^{-3}]$ & $\rm \varepsilon_c [fm^{-4}]$ & $R_{1.4}[{\rm km}]$ & $\Lambda_{1.4}$\\
         \hline
        Case III & 0.50  & 137.5  & 1.91  & 11.78  & 0.88 & 5.14 & 12.46  & 534  \\

        Case II & 0.65  & 132.2  & 2.04  & 12.82  & 0.73 & 4.22 & 13.40  & 1398  \\

        Case IV & 0.70  & 130.6  & 2.10  & 13.25  & 0.70 & 4.04 & 13.86  & 1717 \\ 

        Case I & 0.80  & 127.4  & 2.18  & 13.86  & 0.64 & 3.70 & 14.41  & 2163\\
        \hline
          \multicolumn{7}{c}{MDDQM model Parameters(current work)} \\
          \hline
           Scenario & $\kappa[{\rm MeV^{-1}}]$ & $\sqrt{D}[{\rm MeV}]$ & M$_{\rm max}[{\rm M}_{\odot}]$ & $R[{\rm km}]$ & $\rho_c[{\rm fm}^{-3}]$ & $\rm \varepsilon_c [fm^{-4}]$ & $R_{1.4}[{\rm km}]$ & $\Lambda_{1.4}$\\
         \hline
       Case III & -0.0020  &  126.67  &  1.86  &  12.47  &  0.78  & 4.47 & 13.29  &  1465  \\

       Case II & 0.0003  &  118.82  &  2.10  &  13.20  &  0.68  & 4.26 & 13.74  &  1900  \\ 

        Case I & 0.0031  &  108.54   &  2.30  &  13.75  &  0.60  & 4.05 & 13.92   &  2134\\

        Case IV & 0.0045  &  102.85  &  2.37  &  13.94  &  0.56  & 3.96 & 13.99  &  2189

    \end{tabular}
    \end{ruledtabular}
   
    \label{tabdd}
\end{table*}

\begin{table}[t]
    \centering
        \caption{Mass and radius of the compact stars used as constraints.}
\fontsize{6.6pt}{6.6pt}\selectfont
    \begin{ruledtabular}
    \begin{tabular}{l c c}
    
         \textbf{Star}  &  \textbf{Mass}  &  \textbf{Radius}  \\
         \hline
         PSR J0952$-$0607~\cite{romani2022psr}  &  $2.35 \pm 0.17$M$_{\odot}$  &  ---  \\

         PSR J0740$+$6620~\cite{riley2021nicer}  &   $2.072_{-0.066}^{+0.067}$M$_{\odot}$  &  $12.39_{-0.98}^{+1.30}$km  \\

         PSR J0030$+$0451~\cite{riley2019nicer}  &  $1.34_{-0.16}^{+0.15}$M$_{\odot}$  &  $12.71_{-1.19}^{+1.14}$km  \\ 

         HESS J1731$-$347~\cite{doroshenko2022strangely, abramowski2011new, Sagun:2023rzp}  &  $0.77_{-0.17}^{+0.20}$M$_{\odot}$  &  $10.4_{-0.78}^{+0.86}$km 
    \end{tabular}
    \end{ruledtabular}

    \label{tab1}
\end{table}

As in our previous work~\cite{daSilva:2023okq}, we use Bayesian analysis to optimize the parameters $\sqrt{D}$ and $\kappa$ of the MDDQM model. To compare our current results with the ones of Ref.~\cite{daSilva:2023okq}, we used the same data set of masses and radii of the four compact stars of our previous work, which are shown in Tab.~\ref{tab1}. Like in our previous work~\cite{daSilva:2023okq}, here too, we consider Gaussian likelihood functions and uniform priors. The domain of our priors for $\rm \sqrt{D} = \{0 MeV, 300 MeV\}$ was based on previous works on the DDQM model, as for the prior region for $\kappa= \rm \{-0.01 MeV^{-1}, 0.01 MeV^{-1} \}$ it was obtained assuming small deviations from the value of $\kappa$ that recovers DDQM and the causal limits $\rm M_{max}<3.2 M_{\odot}$ and $R>3M$. We would also like to emphasize that the axis limits in the corner plots were chosen to highlight the regions of highest posterior density, rather than to display the entire prior range. As a result, the plots do not show the full prior ranges. Consequently, any visual proximity of the posterior distribution to the plot boundaries should not be interpreted as the posterior being piled up against the prior limits. Rather, the plots are intentionally zoomed in to improve the visibility and interpretability of the posterior structure, which lies well within the allowed prior space. Here, we optimize the free parameters considering four different cases, namely:

\begin{itemize}
\item {\bf CASE I:} In this case, we searched for the best set of values for $\rm \sqrt{D}$ and $\kappa$ that satisfy the constraints from the two high-mass pulsars, PSR J0952$-$0607 and PSR J0740$+$6620. In this inference, we assumed that the maximum mass could not be smaller than $2.18$M$_{\odot}$; 
 
\item {\bf CASE II:} Here, the focus was to obtain optimized parameters of the MDDQM model that satisfy the constraints imposed by the pulsars, whose masses were precisely measured by NICER, that is, PSR J0030$+$0451 and PSR J0740$+$6620. Here we restricted  $\rm M_{max}$ to be higher than $2.005$M$_{\odot}$;

\item {\bf CASE III:} In this scenario, we optimized the parameters $\rm \sqrt{D}$ and $\kappa$ to describe the low mass compact stars HESS J1731$-$347 and PSR J0030$+$0451. For this case, we assumed $\rm 1.4 \leq M_{\rm max}[M_{\odot}] \leq 2.0$. In this case, we restricted our choice to the points that lead to a decrease $m_i$ at a higher $\rho_B$. This is informed by the pQCD prediction that the effective quark mass decreases with increasing momentum. In this case, the $\rho_B$ is related to spatial momentum, in contrast, in {Cases I and II} this restriction does not lead to any desirable outcome;

\item {\bf Case IV:} Lastly, we look for the best set of values for $\rm \sqrt{D}$ and $\kappa$ that simultaneously satisfy the constraints from all four compact stars. We also assumed that M$_{max}$ could not be smaller than $\rm 2.18 M_{\odot}$.

\end{itemize}
Finally, as a direct constraint in our analysis, we used the mass-radius data for PSR J0740$+$6620 from~\cite{riley2021nicer} and for PSR J0030$+$0451 from~\cite{riley2019nicer}, as these studies reported the smallest radii among groups that analyzed NICER data for these objects with similar \(\rm M_{\rm max}\) values.  We also included PSR J0952$-$0607~\cite{romani2022psr} because it is one of the most massive known NS (\(\sim 2.35 M_{\odot}\)), pushing the upper limits of the EoS stiffness required to support such a mass. There are speculations that it could be a QS, making it particularly relevant for testing EoS models that include deconfined QM.  Lastly, we considered HESS J1731$-$347 due to its unique nature as an extremely low-mass compact object (\(\sim 0.77 M_{\odot}\)). This raises intriguing possibilities: It could be an NS with an unusually soft EoS, or it might be an exotic remnant such as an SQS. Its inclusion helps test whether such a low-mass object could be composed of self-bound QM. These observational constraints, along with theoretical conditions on \(\sqrt{D}\), \(\kappa\), and causality, were used to establish the validity of our EoS models. 
For all four cases, we selected only the values of $\sqrt{D}$ and $\kappa$ that lead to results within the stability window. 

\begin{figure}[t]
\includegraphics[width=0.49\textwidth]{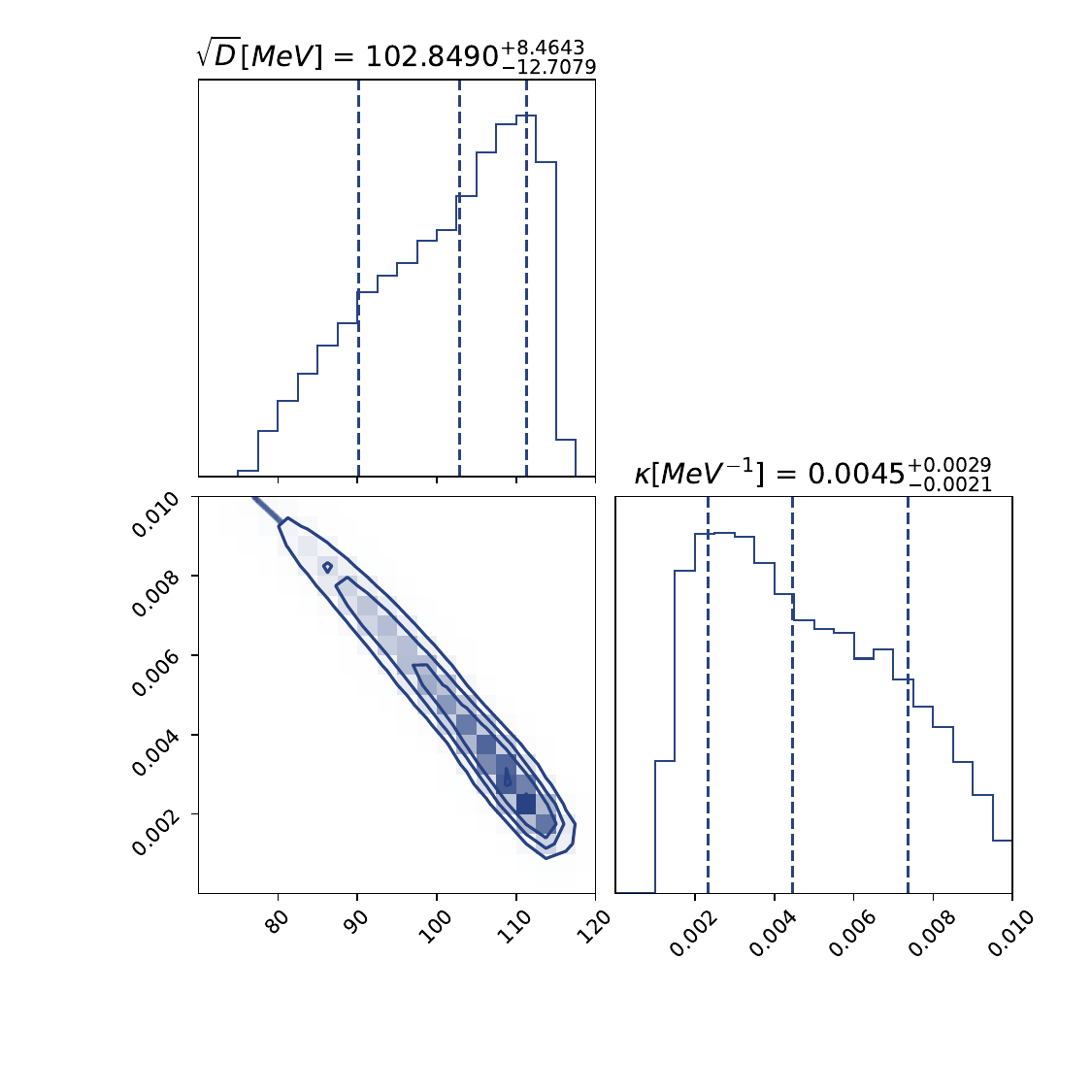}
  \quad
   \includegraphics[width=0.49\textwidth]{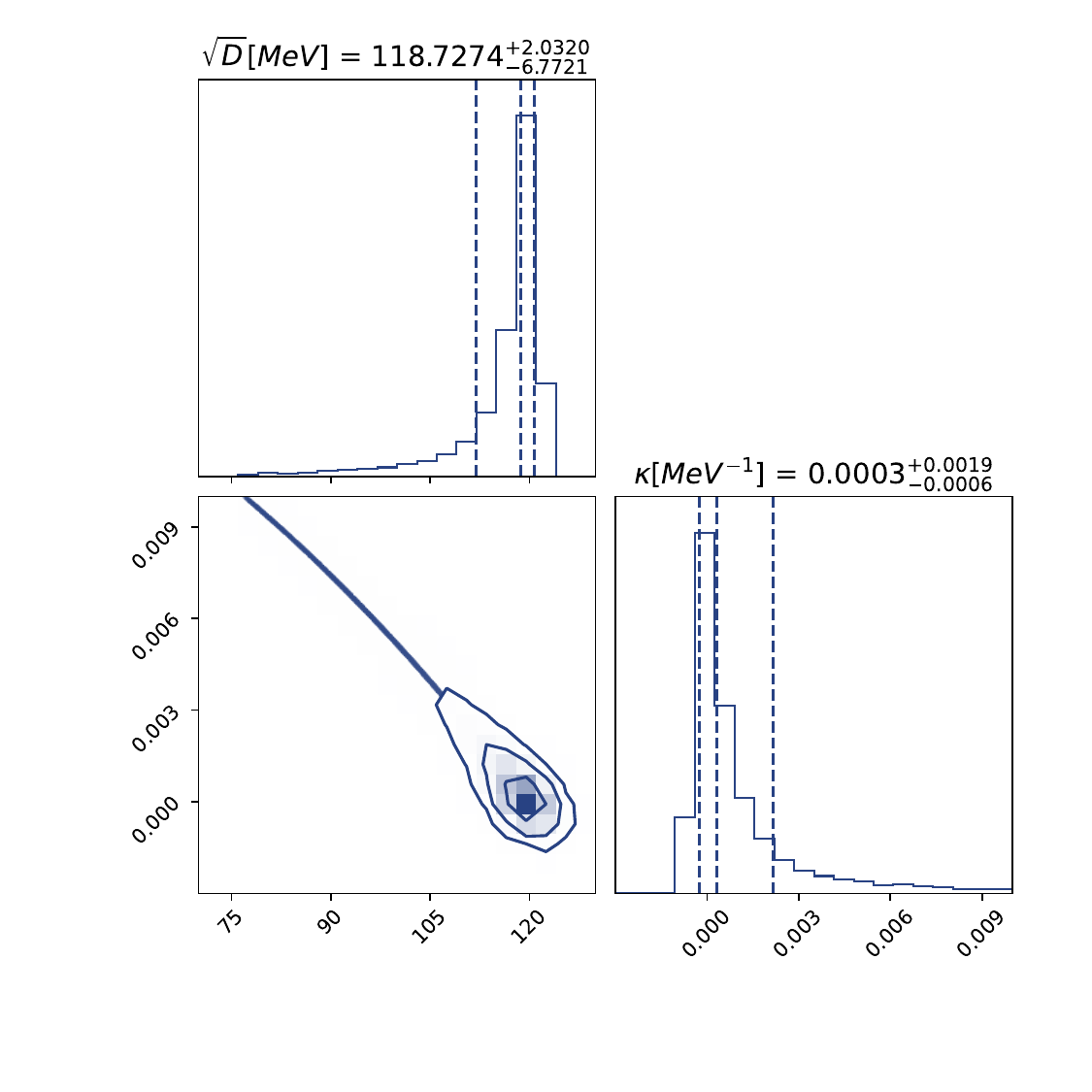}
     
        \caption{Corner plots of the posterior distributions of the parameters $\rm \sqrt{D}$ in MeV and $\kappa$ in MeV$^{-1}$ for the MDDQM model. On the top, we show the results for Case I and on the bottom for Case II.}
         \label{fig1}
\end{figure}

\begin{figure}[t]
    \includegraphics[width=0.49\textwidth]{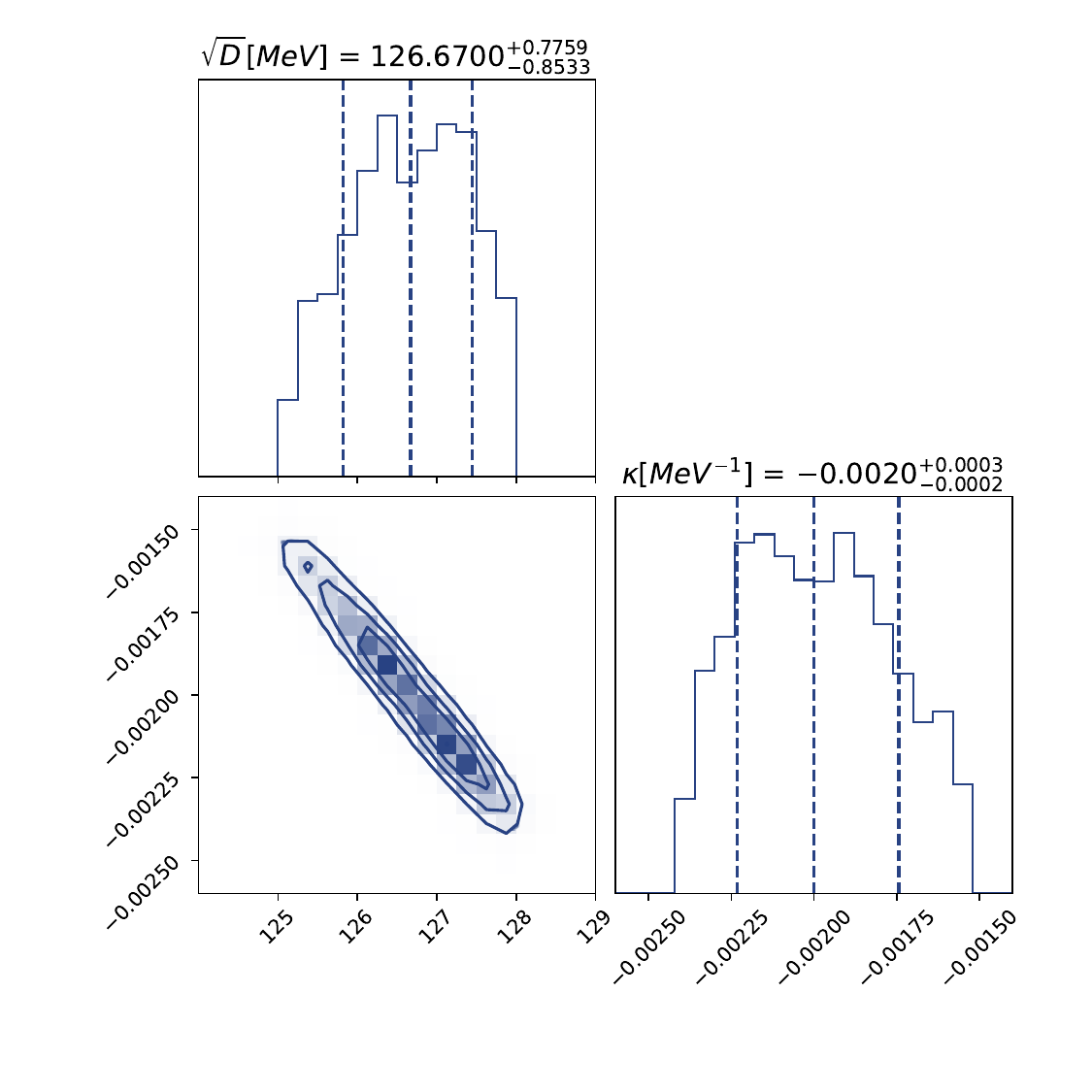}
  \quad
   \includegraphics[width=0.49\textwidth]{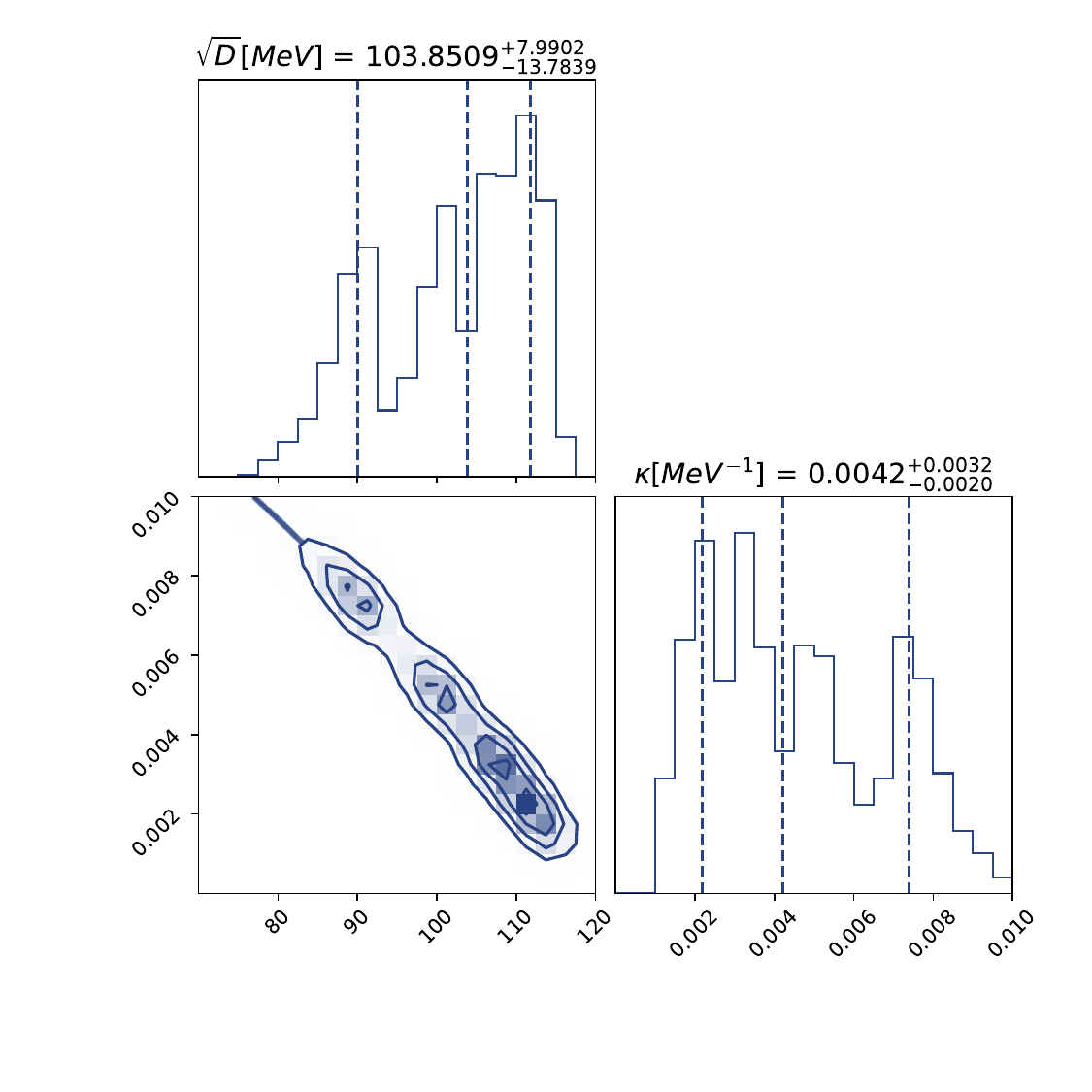}
        \caption{Corner plots of the posterior distributions of the parameters $\rm \sqrt{D}$ in MeV and $\kappa$ in MeV$^{-1}$ for the MDDQM model. On the top, we show the results for Case III, and on the bottom for Case IV.}
         \label{fig2}
\end{figure}

In Fig.~\ref{fig1}, we show the corner plots of the posterior distributions of the parameters of the MDDQM model for Case~I on the top and Case II on the bottom and, in Fig.~\ref{fig2}, we show Case III on the top and case IV on the bottom. In the 1D histograms, the dashed vertical lines denote the 0.16, 0.5, and 0.84 quartiles, and, in the 2D histograms, the contour lines delimit the $\sigma$ levels of each sample of the posterior distributions. By comparing the corner plots for each case, we conclude that if we assume compact stars with small masses, between \(1.4\) and \(2\text{M}_{\odot}\), are strange stars described by the MDDQM model, then the best values for the parameter $\rm \sqrt{D}$ are in the range \([125.82, 127.45]\)MeV, and the best values for \(\kappa\) is in the range \([-0.0022, -0.0017]\)MeV\(^{-1}\). On the opposite side, when we use the MDDQM model to describe stars with high masses, higher than $\rm 2.18 M_{\odot}$, then $\rm \sqrt{D}$ should be in the range $[90.14, 111.31]$MeV and $\kappa$ should be in the range $[0.0024, 0.0074]$MeV$^{-1}$. In Case II, where we use the MDDQM model to describe the data from pulsars PSR J0740$+$6620 and PSR J0030$+$0451, for stellar masses within the ranges $\rm 1.4 M_{\odot}$ and $\rm 2 M_{\odot}$, the suitable values for $\rm \sqrt{D}$ and $\kappa$ are in the region $[111.82, 120.80]$MeV and $[-0.0003, 0.0022]$MeV$^{-1}$, respectively. Lastly, when we look at the corner plot for Case IV (all four stars at the same time) we can observe that it is very similar to the result that we obtained for Case I (only high-mass stars), so we can deduce that the lower limit imposed on $\rm M_{\rm max}$ plays a decisive role in the determination of the best values for the parameters of the MDDQM model.
\begin{figure}[t]
    \includegraphics[width=0.49\textwidth]{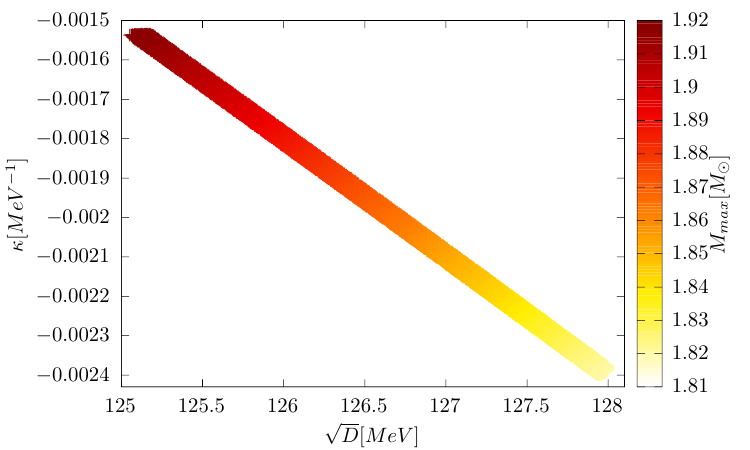}
   \includegraphics[width=0.49\textwidth]{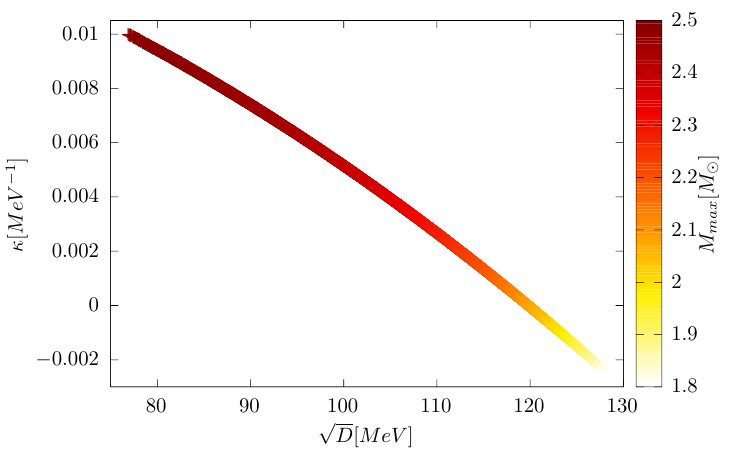}
        \caption{The maximum mass obtained for the values of the parameters $\sqrt{D}$ and $\kappa$ that lie inside the stability window, {\it i.e.}, $(\varepsilon/\rho_B)_{\rm SQM}\leq 930$MeV and $(\varepsilon/\rho_B)_{\rm 2QM} > 930$MeV are satisfied simultaneously. On the top, only the results in which $m_i$ decreases as $\rho_B$ increases are shown, and on the bottom, all results are shown. 
        }
         \label{fig4}
\end{figure}

In Fig.~\ref{fig4}, we show the relation between $\rm M_{max}$ and the parameters $\rm D$ and $\kappa$, for the points that are inside the stability window. In the plot on the top panel, we restrict our analysis to the values of $\rm D$ and $\kappa$ that lead to decreasing $m_i$ with increasing baryon density $\rho_B$. For this case, one can observe that the maximum mass that can be achieved is $\rm 1.92 M_{\odot}$. In the plot at the bottom, we analyze the more general case, where $m_i$ can have any behavior. In this case, the highest value encountered for $\rm M_{max}$ is $\rm 2.5 M_{\odot}$. In both figures, we can readily conclude that the value of $\rm M_{max}$ increases with the increasing $\kappa$ and decreasing $\rm D$.

\begin{figure}[t]
    \includegraphics[scale=1]{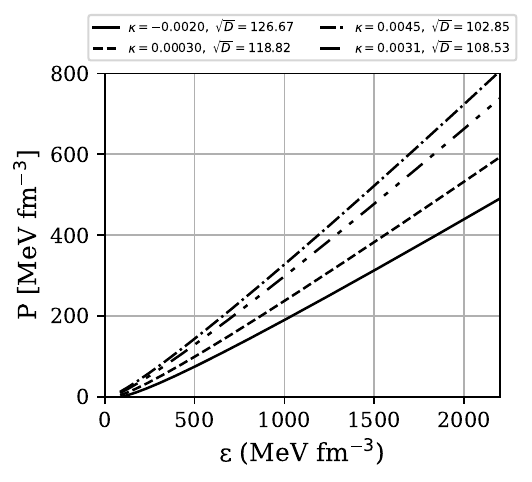}
        \caption{The pressure and energy density are shown respectively on the vertical and horizontal axes. Combining increasing values of $\kappa$ and decreasing values of $\rm \sqrt{D}$, increases the pressure.}
         \label{fig5}
\end{figure}

In Fig.~\ref{fig5}, we show four parameterizations of the MDDQM model to determine how the core pressure of the stars varies with the energy density, one for each of the Cases in which the Bayesian inference was made. For Cases I, II, and III we have chosen the values of $\kappa$ and $\rm \sqrt{D}$ to be the values of the 0.5 quartile. For Case IV, we took the values for $\kappa$ and $\rm \sqrt{D}$ between the 0.16 and 0.84 quartiles that lead to the highest value for the posterior. We chose this because the values of the MDDQM model parameters for the 0.5 quartiles for Cases I and IV are very close, leading to almost identical results. The results show that the EoS is sensitive to $\kappa$, increasing the value of $\kappa$ stiffens the EoS which leads to a considerably enhanced maximum stellar mass. Comparing the value of $\kappa$, negative $\kappa$ generates the least core pressure and inferior $\rm M_{max}$ as shown in Tab.~\ref{tabdd}. The other properties of the QSs that will be discussed subsequently are based on this EoS. 

\begin{figure}
    \centering
    \includegraphics[width=0.5\textwidth]{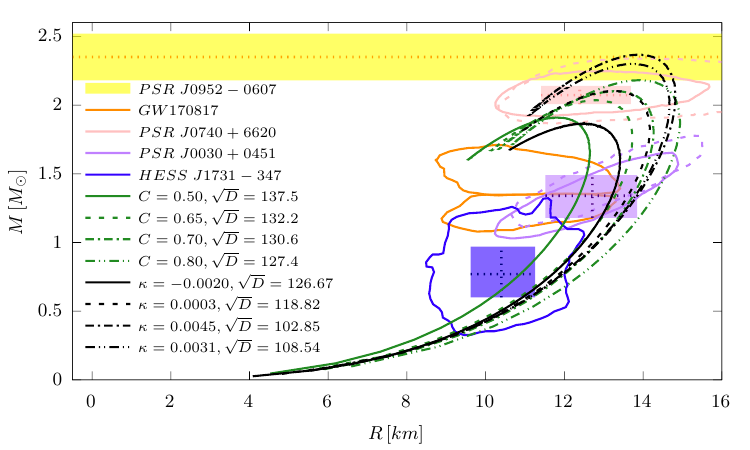}
    \caption{Comparing the mass-radius diagrams obtained from Ref.~\cite{daSilva:2023okq} with the results obtained from modifying the DDQM model. The green lines are the results from the DDQM model, and the black curves are from the MDDQM model. Here $\rm \sqrt{D}$ is in MeV and $\rm \kappa$ is in MeV$^{-1}$ and $C$ is dimensionless.
    }
    \label{fig6}
\end{figure}
In Fig.~\ref{fig6}, we compare the mass-radius diagrams obtained with the EoSs from Fig.~\ref{fig5} for the MDDQM model with the mass-radius diagrams from our previous work~\cite{daSilva:2023okq} on the original DDQM model. It should be noted that aside from the differences in the internal composition of the QSs and NSs, the structure of the QSs is constructed without a crust as discussed in \cite{Kettner1995, melrose2006pair, Melrose:2006zg}. The curves for the latter are in green, and the ones for the modified model are in black, in the same figure. We also show the mass-radius data for HESS J1731$-$347 (dark purple), PSR J0030$+$0451 (light purple), PSR J0740$+$6620 (pink), and PSR J0952$-$0607 (yellow). The graph shows the constraints used in the inference to determine the model parameters in boxes of different colors and the contours of the observed stars. The contours for PSR J0740$+$6620 and PSR J0030$+$0451 are represented in solid and dashed curves for the measurements from the two groups Riley {\it et al.}~\cite{riley2021nicer, riley2019nicer} and Miller {\it et al.}~\cite{Miller:2021qha, Miller:2019cac}, respectively. The first thing we can notice is that with the MDDQM model, we can achieve higher maximum masses than with the  
DDQM model. Besides, the two curves with the highest masses in the modified model, Cases I and IV, have smaller radii than the curve with the highest mass in the DDQM model, hence, they are more compact. However, this decrease in radius is not enough to satisfy the radius constraint from the PSR J0740$+$6620, determined in~\cite{riley2021nicer}. So, the higher masses lead to curves having higher radii. Two groups have determined the radius of PSR J0740$+$6620, 
with marked differences, in~\cite{riley2021nicer} they obtained $R = 12.39^{+ 1.30}_{-0.98}\rm km$ bounded by $16\%$ and $84\%$ quartiles, and in~\cite{Miller:2021qha} they reported $R = 13.71^{+2.61}_{-1.5}\rm km$ at $68\%$ credibility. We used the results reported by the former as the constraint for our analysis since smaller $R$ means highly compact NS.

The curve for Case III has the smallest radii, however, this corresponds to a maximum mass less than the $2$M$_{\odot}$ threshold. The same observation can be made from the 
DDQM model data. The curve for Case II, optimized for the NICER data, satisfies the radius constraint for PSR J0740$+$6620, similar to the curve for $C=0.70$ and $\sqrt{D}= 130.6$ MeV in the DDQM model. Additionally, we observe that the results for the MDDQM model are more favorable for describing HESS J1731$-$347 as a possible strange star since all the analyzed curves satisfy the mass-radius constraint for this small mass compact object, whereas, in the DDQM model, the curve with the highest maximum mass does not meet this constraint.

\begin{figure}
    \centering
    \includegraphics[width=0.5\textwidth]{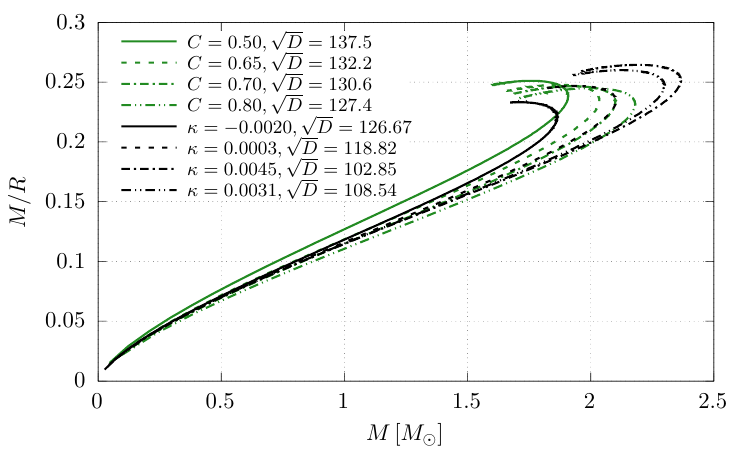}
    \caption{Comparing the compactness as a function of the mass obtained from Ref.~\cite{daSilva:2023okq} with the results obtained from modifying the DDQM model. The green lines are the results from the DDQM model, and the black curves are from the MDDQM model. Here $\rm \sqrt{D}$ is in MeV, $\kappa$ is in MeV$^{-1}$ and $C$ is dimensionless.}
    \label{fig7}
\end{figure}

The compactness $M/R$ as a function of the mass is shown in Fig.~\ref{fig7} for the DDQM model in green and the modified model in black. We observe that for QSs with masses less than $\sim 1.9\text{M}_{\odot}$, the curve for $C=0.50$ and $\rm \sqrt{D}=137.5$MeV (Case III of DDQM), in the DDQM model is more compact than its counterpart from the modified model. In Fig.~\ref{fig6}, we find that the curves for Case II of MDDQM and  $C=0.70$ and $\rm \sqrt{D}=130.6$MeV, which correspond to Case IV in the DDQM model, have similar $\rm M_{max}$ and $R$. In Fig.~\ref{fig7}, we observe that the compactness for these two parameterizations is similar, compared to the others. In the cases of higher maximum mass, we observe that the curves for Cases I and IV in the modified model produce more compact stars than the ones from $C=0.80$ and $\rm \sqrt{D}=127.4$MeV (Case I of DDQM) obtained from our previous work, which yielded the highest $\rm M_{max}$. Also, the curve for Case I of MDDQM is the one that reached the highest value of compactness from all the cases analyzed. In effect, the modified model leads to highly compact QSs compared to the DDQM model.
\begin{figure}
    \centering
    \includegraphics[width=.5\textwidth]{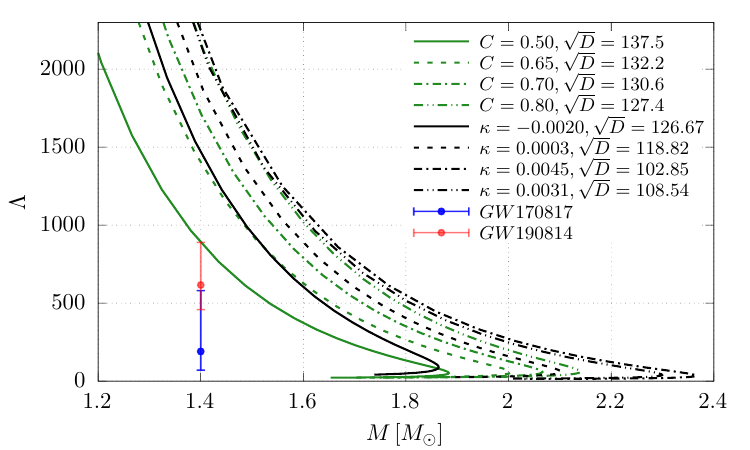}
    \caption{Comparing the dimensionless tidal deformability as a function of the mass obtained from Ref.~\cite{daSilva:2023okq} with the results obtained from modifying the DDQM model. The green lines are the results from the DDQM model and the black curves are from the MDDQM model. Here $\rm \sqrt{D}$ is in MeV and $\kappa$ is in MeV$^{-1}$ and $C$ is dimensionless.
    }
    \label{fig8}
\end{figure}
Another astrophysical quantity of interest, the dimensionless tidal deformability, $\Lambda$, as a function of stellar mass, is shown in Fig.~\ref{fig8}, with the curves of the MDDQM model in black and the curves from our previous work in green. We can observe that, in general, the strange stars obtained in the DDQM model have a smaller deformability than the ones obtained in the modified model. In addition, it is interesting to notice that the EoS parameterizations that lead to similar mass-radius and compactness-mass relations, as is the case for the EoSs for Case II of MDDQM $(\kappa=0.0003,\sqrt{D}=118.82$MeV) and for Case IV of DDQM $(C=0.70, \sqrt{D}=130.6$MeV), can present noticeable differences in their tidal deformabilities, due to the differences in the energy density discontinuity at the surface of each QS. 
\begin{figure*}[t]
    \includegraphics[scale=0.8]{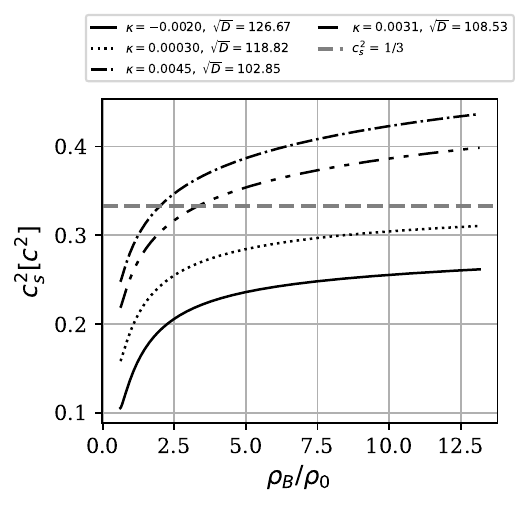}
    \includegraphics[scale=0.8]{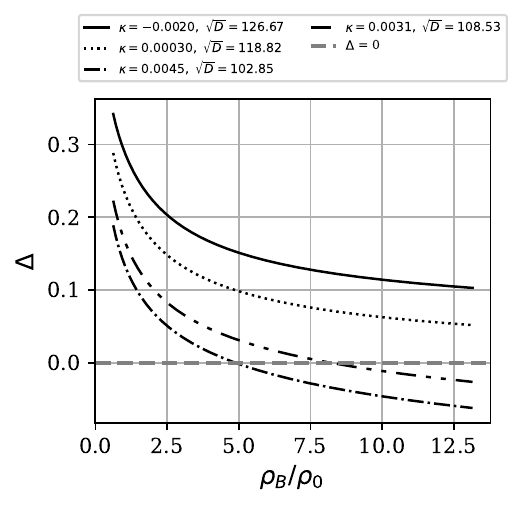}
     \includegraphics[scale=0.8]{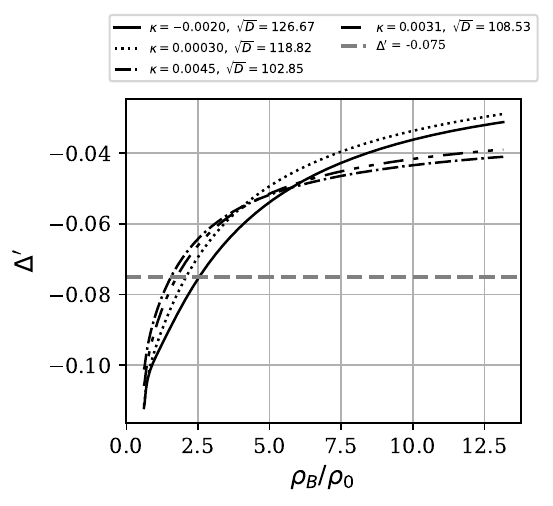}
      \includegraphics[scale=0.8]{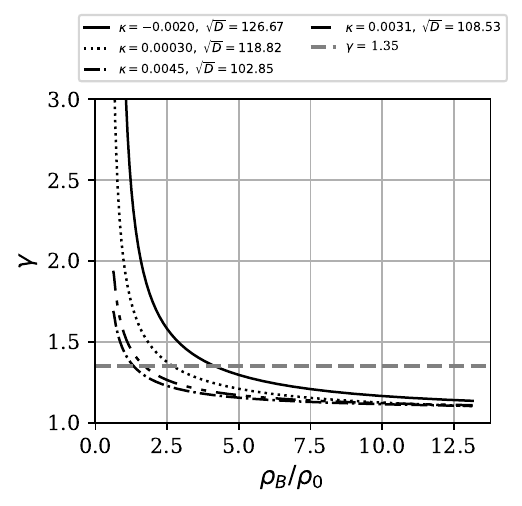}
       \includegraphics[scale=0.8]{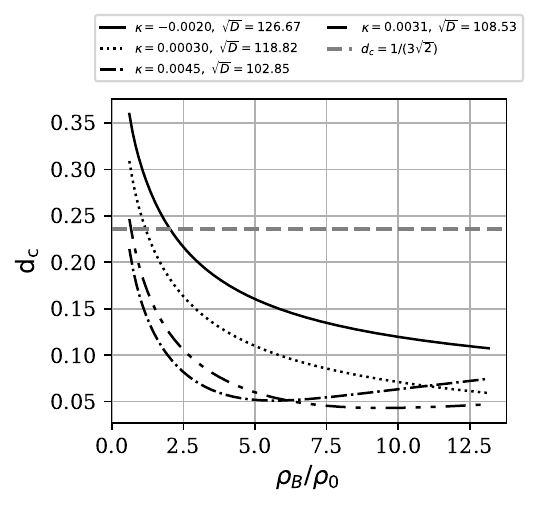}
       \includegraphics[scale=0.8]{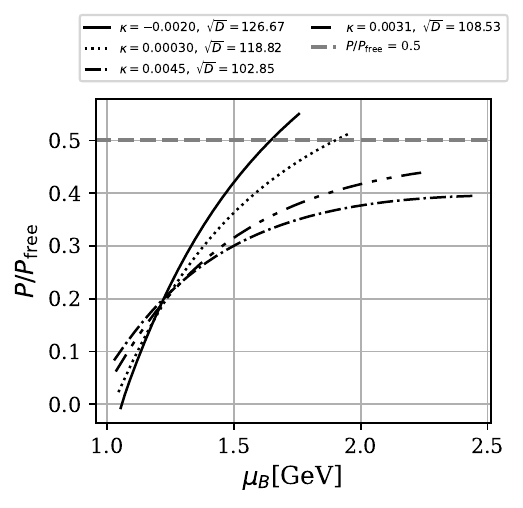}
    
        \caption{In this figure, we show the behavior of six dimensionless quantities presented in Tab.~\ref{tab1a}, using either the conformal or perturbative limit as a benchmark in most cases for our analysis. We show $c_s^2, \, \gamma,\,\Delta,\, d_c,\, {\rm and}\, \Delta'$ as a function of $\rho_B/\rho_0$ and the pressure of the QM normalized by the free non-interacting Fermi-Dirac pressure, $P/P_{\rm free}$, as a function of $\mu_B$. The gray line in each figure was chosen based on the following considerations: the line in $c^2_s$ was chosen using the conformal limit, and in $\gamma$ the limit was placed at the average of the values obtained from perturbative QM because other authors have obtained a higher value up to $\gamma = 1.75$ (see, \textit{e.g.},~\cite{Annala:2019puf}) as the possible conformal limit. Additionally, the conformal limit was chosen for $\Delta$, the lower limit of the prediction of pQCD was chosen for $P/P_{\rm free.}$, and an average value in the pQCD data was taken for $\Delta'$, also, the pQCD value was used for the $\rm d_c$ threshold.
       }
         \label{fig9}
\end{figure*}

Before we discuss Fig.~\ref{fig9} in detail, we would like to highlight the characteristics of the $c^2_s$ at various densities and how they affect our results in this section. The $c_s^2$ is necessary for astrophysical applications because it relates to the stiffness of the EoS. In the low-density region, $\sim 2\rho_B$ where CEFT can effectively describe {hadronic} matter with pion and nucleon effective degrees of freedom, $c_s^2$ is known to show a rapid increase with $\rho_B$~\cite{Leonhardt:2019fua}.  Studies that constrain NS masses point to the existence of a maximum $c_s^2$ at densities $\rho_B \lesssim 10\rho_0$ that far exceeds the asymptotic value in a non-interacting quark gas~\cite{Huth:2020ozf, Gorda:2022jvk, Braun:2022jme}. The perturbative QCD predicts that the conformal limit, $c_s^2=1/3$, is approached from below~\cite{Bedaque:2014sqa, Contrera:2022tqh, Traversi:2021fad} in a dense matter medium and saturates in exactly conformal matter at very high densities reachable only by pQCD. So we make deductions from the characteristics of the stellar matter through the behavior of $c_s^2$ and other quantities such as $\Delta$, and $\gamma$ based on how fast the conformal limit is approached from below or violated towards the high-density regions. The $c_s^2$ has been used as a benchmark to investigate the near-conformality in NS matter in~\cite{Fujimoto:2022ohj, Marczenko:2022jhl, Bedaque:2014sqa}, hybrid NSs in~\cite{Annala:2019puf, Annala:2023cwx}, and QSs~\cite{Albino:2021zml, daSilva:2023okq} at densities well within the NS densities ($5-10)\rho_0$.

 In Fig.~\ref{fig9}, we expect the characteristics of EoSs for the QSs to approach the near-conformal behavior predicted to exist among strongly interacting QM at high densities. 
  However, the results show that in sufficiently larger QSs, the near-conformality expected in QM is strongly violated. As can be clearly seen from $c_s^2(\rho_B/\rho_0),\, \Delta(\rho_B/\rho_0),\, {\rm and}\; P/P_{\rm free}(\mu_B)$ curves. Stars with $\rm M > 2.10 M_\odot$, violate the conformality criteria towards the higher $\rho_B$ regions. We also observe a violation of this conformality criterion in other models for SQSs that reach the $\rm M>2.10M_\odot$ threshold; for example, in the MIT bag model~\cite{Lopes_2021b, daSilva:2023okq}, we observe $c_s^2>1/3$.  These violations are attributed to multi-quark states and, presumably, high gluon densities. Since higher maximum gravitational mass means higher pressure within the star, high pressure favors multi-quark generation and, consequently, easy formation of condensates from diquarks and quark-antiquark pairs. It is important to note that the behavior of the QM for all EoSs falls within the range of DNM and pQCD presented in Tab.~\ref{tab1a}. On the contrary, the conformal limit set through the CFT was not approached from below, as expected in some cases. However, different values for near-conformal limits in dense matter have been determined in~\cite{Annala:2019puf, Annala:2023cwx}, which are considerably different from the values in Tab.~\ref{tab1a}.

The curve for $P/P_{\rm free}(\mu_B)$~\cite{Kurkela:2009gj, Gorda:2018gpy} tells us about the possible degrees of freedom of the quarks and the gluons of the matter. The EoSs that cross the dashed gray line in the last figure in Fig.~\ref{fig9} have higher degrees of freedom than those below it. Therefore, near-conformal QM EoSs are expected to cross the dashed gray line, while the EoSs that violate the near-conformability are expected to fall below the dashed gray line. This classification is based on the degrees of freedom of the matter, where a deconfined QM has a larger number of degrees of freedom than confined matter. Likewise, the two EoSs or Cases I and IV of MDDQM ($\rm \sqrt{D}=108.54$MeV, $\kappa = 0.0031$MeV$^{-1}$ and $\rm \sqrt{D}=102.85$MeV, $\kappa = 0.0045$MeV$^{-1}$, respectively) violate the near-conformality established through $c_s^2(\rho_B/\rho_0)$, ($c_s^2 \leq 1/3$)~\cite{Cherman:2009tw} and $\Delta(\rho_B/\rho_0)$, ($\Delta \rightarrow 0$)~\cite{Fujimoto:2022ohj}. On the other hand, all the EoSs satisfy the prediction of pQCD for $\gamma$~\cite{Kurkela:2009gj} while approaching the conformal limit above, $\gamma > 1$. The small negative value required for $\Delta'$, in the pQCD case, was satisfied by the EoSs of Cases III and II of MDDQM, which have the two relatively lighter stars ($\rm \sqrt{D}=126.67$MeV, $\kappa = -0.0020$MeV$^{-1}$ and $\rm \sqrt{D}=118.82$MeV, $\kappa = 0.0003$MeV$^{-1}$, respectively) attaining smaller negative values at higher $\rho_B$. In the case of $d_c$, all EoS fall within the pQCD limit at higher $\rho_B$. From these results, we can infer that no single quantity is sufficient to determine and classify approximately conformal matter.  

From the plots in Fig.~\ref{fig9}, $c_s$, $\Delta$ and $P/P_{\rm free}$ show that the QM determined from the EoSs of Cases III and II of MDDQM, $\rm \sqrt{D}=126.67$MeV, $\kappa = -0.0020$MeV$^{-1}$ and $\rm \sqrt{D}=118.82$MeV, $\kappa = 0.0003$MeV$^{-1}$ show near-conformal characteristic approaching $c_s^2 =1/3$ from below, crossing $\Delta$ towards the negative regions, and also crossing the gray line in $P/P_{\rm free}$ towards the strong positive value. Aside from satisfying DNM and pQCD limits at higher $\rho_B$, contrary to the EoSs of Cases IV and I of MDDQM, $\rm \sqrt{D}=108.54$MeV, $\kappa = 0.0031$MeV$^{-1}$ and $\rm \sqrt{D}=102.85$MeV, $\kappa = 0.0045$MeV$^{-1}$, which {satisfy the DNM and pQCD predictions but not CFT predictions based on the data in Tab.~\ref{tab1a}}, furthermore, the negative value of $\Delta$ in the core of sufficiently massive stars implies that towards the core $P>\varepsilon$, for the two relatively small stars. It is worth emphasizing that for exactly conformal matter, the $\Delta$ saturates at $\Delta =0$. A positive QCD trace anomaly has been predicted in the literature at finite temperature~\cite{PhysRevD.27.140}, lattice QCD predicts a similar outcome~\cite{HotQCD:2014kol, Borsanyi:2013bia}, and other phenomenological nuclear matter EoS models predict a negative trace anomaly due to sudden stiffening of the EoS~\cite{Serot:1997xg, Akmal:1998cf}. From the discussions above to determine the {near-conformality} in QM, other properties of the QM need to be studied to reach a firm conclusion. Analysing the central baryon densities in Tab.~\ref{tabdd}, we observe that QSs with smaller $\rm M_{max}$ in both the DDQM model and MDDQM model are associated with higher values of $\rho_c$. We find a similar trend of $\rho_c$ values in~\cite{daSilva:2023okq}, where the authors used the vector MIT bag model to study QSs. 

In SQSs, a larger mass corresponds to a smaller \(\rho_c\), which results in a more compact star. This contrasts with typical NSs, where higher masses require higher central densities to support the additional gravitational pressure. The EoS of QM differs significantly from that of nuclear matter in NSs, both in terms of composition and the mode of interaction between particles. QM allows for denser stellar configurations because the deconfined quarks are more tightly packed, which affects the star's radius. The advantage of a decreasing $\rho_c$ with increasing mass in QSs is that it enables greater stability, higher maximum masses, and more compact stellar configurations than in NSs due to the unique properties of QM \cite{Glendenning2012}. A similar trend is observed for the central energy density, $\varepsilon_c$. The SQSs with $\rm M_{\rm max}<2M_\odot$ in the DDQM and the MDDQM models have higher $\rho_c$, thus, we can infer that stars with higher $\rho_c$ are more likely to behave as if the quarks are in a deconfined state. 

\begin{figure}[t]
    \includegraphics[scale=1]{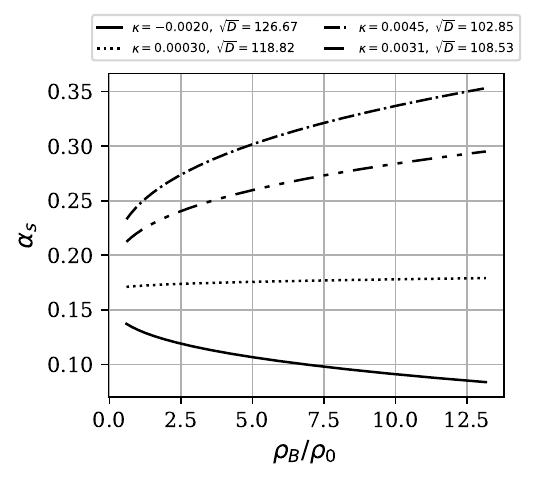}
    \includegraphics[scale=1]{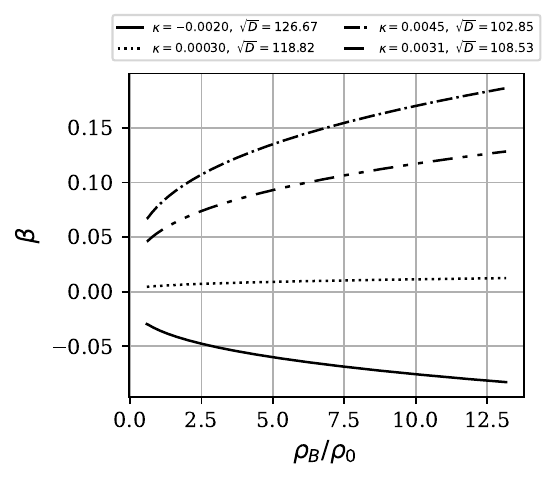}
        \caption{The effective running coupling constant (top panel) and the $\beta$-function (bottom panel) as a function of $\rho_B$ for different values of $\kappa$.
        }
         \label{fig10}
\end{figure}

In Fig.~\ref{fig10}, one can see that the curves are sensitive to the sign of $\kappa$. When $\kappa<0$ as observed in $\rm \sqrt{D}=126.67 MeV, \kappa = -0.0020 MeV^{-1}$ {\color{blue} (Case III)}, the $\alpha_s(\rho_B)$ and $\beta$-function curves behave like the ones predicted by the QCD theory~\cite{Deur:2016tte}, \textit{i.e.}, the $\alpha_s$ and the $\beta$-function decrease with $\rho_B$, the quarks gradually become free as $\rho_B$ increases, and this choice produces a negative $\beta$-function as expected. The data obtained with these parameters generate a QS, whose interior will present completely deconfined quarks at $\rho_B \approx 126.51\rho_0$ when we set $m_i-m_{i0} =0$ in Eq.~(\ref{a1}) and determine $\rho_B$ (the exact $\rho_B$ at which QM becomes completely deconfined is not yet known in the literature; in contrast, some authors have estimated $\rho_B > 40\rho_0$~\cite{Kurkela:2009gj}, while in other conservative models, they estimate $\rho_B\gtrsim(10-40)\rho_0$). However, it is important to mention that the maximum stellar mass obtained from this choice is lower than the $\rm 2M_\odot$ threshold required for NSs, determined through observation. When $\kappa$ increases to $\kappa = 0.0003$MeV$^{-1}$, the maximum stellar mass reaches $2.10 \rm M_\odot$ but the $\alpha_s$ and $\beta$-function curves show an approximately constant behavior which is not the usual decreasing $\alpha_s$ with $\rho_B$ and negative $\beta$-function expected. Qualitatively, it shows a deviation from the known QCD behavior but satisfies other {near-conformal} characteristics of the quantities presented in Tab.~\ref{tab1a}. 

That notwithstanding, looking at the corner plot in Case II of Fig.~\ref{fig1}, some of the suitable values of $\kappa$ fall within the range of negative values, which yields the correct behavior of the $\beta$-function (see Eq.~(\ref{beta})). Still, they lie within the lowest percentile region. As $\kappa$ becomes positive and $\rm M_{max}$ increases beyond $2.10 \rm M_\odot$, the behavior of $\alpha_s$, and $\beta$-function strongly deviates from the QCD predictions. The $\alpha_s$ and $\beta$-function increase monotonically with $\rho_B$ towards a stronger positive value. Similar deviation is observed in Fig.~\ref{fig9}, where $\rm \sqrt{D}=108.54$MeV, $\kappa = 0.0031$MeV$^{-1}$ and $\rm \sqrt{D}=102.85$MeV, $\kappa = 0.0045$MeV$^{-1}$  (Cases IV and I) show strong violation to the conformality criteria of the CFT. Nonetheless, the $\kappa> 0$ choice generally satisfies the required $\rm 2M_\odot$ threshold. Comparing Figs.~\ref{fig9} ($\rm c_s^2,\, \Delta,\, and\, P/P_{free}$) and~\ref{fig10}, we can say that the QM is perturbative, dense, and shows near-conformal characteristics for $\rm \sqrt{D}=126.67$MeV, $\kappa = -0.0020$MeV$^{-1}$ and $\rm \sqrt{D}=118.82$MeV, $\kappa = 0.0003$MeV$^{-1}$ (Cases III and II) parameterizations.

\section{Final Remarks}\label{fr}
The results of this work are twofold. First, we determine the microscopic structure of the star by imposing astrophysical constraints and then use the resulting model parameters to infer the microscopic properties of the stellar matter. The direct outcomes of the Bayesian analysis of the stellar structure include the mass-radius diagram, compactness, and tidal deformability. The model parameters derived from this analysis are subsequently used to assess the near-conformality, or otherwise, of the QM.

The MDDQM model led to higher maximum stellar masses and greater compactness than the commonly used model. Such modifications have become necessary due to the recent discoveries of supermassive NSs that rule out QSs with smaller gravitational masses~\cite{ozel2006soft,rodrigues2010quark, Albino:2021zml} as possible candidates for NSs. Aside from that, we determine the model parameters in an optimized manner and investigate the strong interacting characteristics of the QM relative to increasing stellar mass. Here, we compute six different quantities that help to study the near-conformality or otherwise of the QM besides the effective running coupling constant and the $\beta$-function, whose behavior is well established in QCD. We observe that the QM that composes heavier QSs such as 2.30  and $\rm 2.37 M_\odot$, behaves as though the stars are composed of QM in a confined state, contrary to the near-conformal behavior expected in high-density matter. We attribute this unexpected behavior of the QM at higher $\rho_B$ to the formation of presumably multi-quark states and color glass condensates, influenced by the pressure build-up in the stellar core as its $\rm M_{max}$ increases, due to the repulsive interaction term in the model. It has also been shown in~\cite{rodrigues2010quark, Albino:2021zml,alford2007quark,franzon2012self,klahn2015vector, Lopes_2021,song2019effective,otto2020nonperturbative} that the presence of repulsive interactions in quark matter models leads to high stellar masses similar to what we observed. We also observe that the QM that composes QSs with $\rm M_{max}$'s between $\rm 1.86 M_\odot$ and $2.10 \rm M_\odot$ behave as if they are in near-conformal state, largely satisfying pQCD, DNM, and CFT, predictions at higher $\rho_B$ as shown in Tab.~\ref{tab1a} and Fig.~\ref{fig9}. The well-known $\alpha_s$ and the $\beta$-function determined in Fig.~\ref{fig10} show the desired behavior for $\rm M \leq 2.10 M_\odot$.

Our work establishes within the MDDQM model framework that the QM with $\kappa > 0$ parameterizations produces heavier QSs but violates the CFT criteria. On the other hand, the QM that composes less massive QSs, with $\kappa < 0$ parameterizations, satisfies the near-conformal criteria at high-density regions. We calculate the stellar properties such as the mass-radius diagram, compactness, and tidal deformability, in Figs.~\ref{fig6},~\ref{fig7}, and~\ref{fig8} respectively, through the EoSs, Fig.~\ref{fig5}. As expected, the stiffer EoSs yielded higher $\rm M_{max}$, high $M/R$, and higher $\Lambda_{1.4}$ than the softer ones, accordingly. The modification was intended to lead to stars with enhanced stellar properties than the original DDQM model, so we compare the results with the ones in~\cite{daSilva:2023okq} to establish the difference. The modified model produces QSs with enhanced mass-radius and compactness characteristics as intended for the same astrophysical constraints as the DDQM model.

Comparing the tidal deformability of the two models in Tab.~\ref{tabdd}, we observe that the modified model produces larger $\Lambda_{1.4}$ than the ones in~\cite{daSilva:2023okq} in all cases. Still, none of the two satisfies the measured $\Lambda_{1.4}$ of GW170817~\cite{LIGOScientific:2017vwq}. On the other hand, the DDQM model mildly satisfies the $\Lambda_{1.4}$ at the upper limit of the secondary component of the GW190814 event~\cite{LIGOScientific:2020zkf}, assuming that the secondary component can be described as a massive compact star. Even though there is an unsettled debate on the nature of this mass gap object, with other researchers believing that it could be an NS on one hand, and others believing it is a black hole on the other hand, the possibility of it being SQS has also been discussed in~\cite{Oikonomou:2023otn, Bombaci:2020vgw}. Hence, both models violate the $\Lambda_{1.4}$ for binary NS mergers, which coincides with the predictions of other phenomenological quark matter models, such as the confining quark model~\cite{Cao:2020zxi}, quasiparticle model~\cite{ Chu:2021aaz}, and a quarkyonic matter model~\cite{Zhao:2020dvu}, among others. As a result, the binary NS merger that led to the observed GW170817 event is unlikely to be composed of QSs.

Also, comparing the central baryon densities of the DDQM model and the MDDQM model presented in Tab.~\ref{tabdd}, we find that the $\rho_c$ of the MDDQM model generally shifts towards lower values of $\rho_c$. Additionally, lower mass QSs are associated with higher $\rho_c$ values, for instance, the lightest QS in the MDDQM model framework, $1.86\rm M_\odot$, is associated with $\rho_c =0.78\rm  fm^{-3}$ compared to the heavier QS, $2.37\rm M_\odot$, with $\rho_c = 0.56\rm  fm^{-3}$. Consequently, the QM behaves in a near-conformal manner when $\rho_c$ is higher. The trend of increasing $\rho_c$ with decreasing QS mass was also reported in~\cite{daSilva:2023okq}, where the authors analyzed QSs built from the DDQM model and the vector MIT bag model.

The specific findings are summarized below:
\begin{itemize}
    \item The model parameters were fixed using Bayesian inference to compare the results with the ones determined in~\cite{daSilva:2023okq} using the 
    DDQM model. Both models comprised two free model parameters that needed to be fixed. The corner plots of the posterior distributions are presented in Figs.~\ref{fig1} and~\ref{fig2}. The relation between the free parameters linked to the stellar masses was also shown in Fig.~\ref{fig4}.

    \item The EoSs shown in Fig.~\ref{fig5} and model parameters in Fig.~\ref{fig4} demonstrate that the pressure in the stellar core increases with increasing $\kappa$ and decreasing $D$. The increase in $\kappa$ is also reflected in the mass-radius diagram in Fig.~\ref{fig6}, where stiffer EoS corresponds to higher $\rm M_{max}$.

    \item In Fig.~\ref{fig7}, we observed that the $\kappa<0$, produces a less compact QSs with larger radii and smaller $\rm M_{max}$ compared to $\kappa>0$. Comparing the results with the ones obtained in~\cite{daSilva:2023okq}, we observed that aside from $\kappa<0$ case, our model produces QSs with enhanced $\rm M_{max}$ and compactness. The main weaknesses of both models (DDQM model and MDDQM models) are that they produce large $\rm \Lambda_{1.4}$ due to higher radii ($\rm R>13km$) and higher $\Delta\varepsilon$ as presented in Fig.~\ref{fig8}. However, obtaining higher compactness and $\rm M_{max}$, in line with our current objective, requires a stiffer EoS at higher densities, while achieving the required value of $\Lambda_{1.4}$ in the GW170817 event necessitates a softer EoS at lower densities to reduce tidal deformability \cite{LIGOScientific:2018cki}. We intend to address this weakness in the future through further modifications toward achieving a unified EoS for QSs.

    \item We explore the near-conformal characteristics of the QM towards higher $\rho_B$ by studying various quantities in Tab.~\ref{tab1a} since no individual quantity is a sufficient condition for classifying conformal matter behavior in NSs. The results in Fig.~\ref{fig9} demonstrate that $\rm \sqrt{D}=126.67$MeV, $\kappa = -0.0020$MeV$^{-1}$ and $\rm \sqrt{D}=118.82$MeV, $\kappa = 0.0003$MeV$^{-1}$ (Cases III and II) parameterization points to QSs made up of {near-conformal QM} while $\rm \sqrt{D}=108.54$MeV, $\kappa = 0.0031$MeV$^{-1}$ and $\rm \sqrt{D}=102.85$MeV, $\kappa = 0.0045$MeV$^{-1}$ (Cases IV and I) points to QSs made up of matter that violates the near-conformality threshold expected in dense QM at higher densities. Out of the two parameters that conform with the near-conformality analyzed in Fig.~\ref{fig9}, $\rm \sqrt{D}=118.82$MeV, $\kappa = 0.0003$MeV$^{-1}$, leads to a QS with $M > 2\,\rm M_\odot$. Extending the analysis to the effective running coupling and the $\beta$-function, investigated as a function of the $\rho_B$, and the findings presented in Fig.~\ref{fig10}, we can affirm that $\rm \sqrt{D}=126.67$MeV, $\kappa = -0.0020$MeV$^{-1}$ (Case III) parameterization leads to QSs composed of near-conformal QM in its interior. 
\end{itemize}
The current work aims to provide insight into the long-standing problem of the properties of dense QM at various densities by employing the QS model, which allows us to investigate the behavior of QM under extreme density conditions comparable to core densities of massive NSs. From this work, we have established that even though QM is generally expected to be approximately conformal, not all the QMs that compose QSs are near-conformal. Indeed, some QSs may exist as if they are composed of strongly bound QM, as the model reveals. This observation challenges the current understanding of the possible forms of quark cores in hybrid NSs. A natural extension of this work is to look at the formation and the behavior of quark cores in massive NSs.

\section*{Acknowledgements}
A.I. would like to thank the S\~ao Paulo State Research Foundation (FAPESP) for financial support through Grant No.  2023/09545-1. T. F. thanks
the partial financial support from the Brazilian Institutions: Conselho Nacional de Desenvolvimento Cient\'ifico e Tecnol\'ogico (CNPq) (Grant No. 306834/2022-7), Coordena\c{c}\~ao de Aperfei\c{c}oamento de Pessoal de N\'ivel Superior (CAPES) (Finance Code 001) and FAPESP (Grant 2019/07767-1). This work is a part of the
project Instituto Nacional de  Ci\^{e}ncia e Tecnologia - F\'isica
Nuclear e Aplica\c{c}\~{o}es  Proc. No. 464898/2014-5. Special thanks to the Laborat\'orio Multiusu\'ario de Pesquisas F\'isicas (LAMPEF) for providing the cluster infrastructure used in the Bayesian inferences. L.C.N.S would like to thank FAPESC for financial support under grant 735/2024
and D.P.M. is partially supported by CNPq under Grant No. 303490/2021-7.

\section{Appendix}
\addcontentsline{toc}{section}{Appendix}

\subsection{Thermodynamic Consistency}\label{ap1}
The Helmholtz free energy \(f\) as a function of \(m_i\) and \(\mu_i^*\) is  given by:
\begin{equation}
    f = \Omega_0(\{\mu_i^*\}, \{m_i\}) + \sum_i \mu_i^* \rho_i,
\end{equation} 
where \(\Omega_0\) is the thermodynamic potential and \(\rho_i\) is the number density given by  
\begin{equation}
    \rho_i = \frac{g_i k_{fi}^3}{6\pi^2},
\end{equation}
and the total baryon density is 
\begin{equation}
    \rho_B = \frac{1}{3} \sum_i \rho_i.
\end{equation}
 
The relation between \(\mu_i^*\) and \(k_{fi}\) is 
\begin{equation}
    k_{fi} = \sqrt{\mu_i^{*2} - m_i^2}, \qquad \mu_i^* = \sqrt{k_{fi}^2 + m_i^2},
\end{equation} 
with the relationship between \(\mu_i\) and \(\mu_i^*\) given by  
\begin{equation}
 \mu_i = \mu_i^* - \mu_I,   
\end{equation}
where \(\mu_I\) is the interaction term. Although modifying \(m_i\) and \(\mu_i\) does not change the form of \(\rho_i\), it alters pressure \(P\) and energy density \(\varepsilon\) as shown in~\cite{peng2008deconfinement, Peng:2000ff}. The thermodynamic potential \(\Omega_0\) is  
\begin{align}
    \Omega_0(\{\rho_i\}, \{m_i\}) &= -\sum_i \frac{g_i}{24\pi^2}\Big[\mu_i^* k_{fi} \left(k_{fi}^2 - \frac{3}{2} m_i^2\right) \nonumber\\&+ \frac{3}{2} m_i^4 \ln\frac{\mu_i^* + k_{fi}}{m_i}\Big],
\end{align}
  
leading to  
\begin{equation}
    P = -\Omega_0 + \sum_{i,j} \rho_i \frac{\partial \Omega_0}{\partial m_j} \frac{\partial m_j}{\partial \rho_i},
\end{equation}
and  
\begin{equation}
    \varepsilon = \Omega_0 + \sum_i \mu_i^* \rho_i.
\end{equation}

\subsection{TOV Equations}\label{appendix:TOV}

The TOV equations are given by:
\begin{align}
    \frac{dP(r)}{dr}&=-[\varepsilon(r) + P(r)]\frac{M(r)+4\pi r^3 P(r)}{r^2-2M(r)r}, \label{eqq1} \\
    \frac{dM(r)}{dr}&=4\pi r^2 \varepsilon(r), \label{eqq2}
\end{align}
where $M(r)$ is the gravitational mass of a spherically symmetric compact star, in these equations, $P(r)$ represents the pressure, and $\varepsilon(r)$ denotes the energy density. We have adopted natural units where $G = c = 1$. For realistic EoSs, Eqs.~(\ref{eqq1}) and~(\ref{eqq2}) typically require numerical techniques to solve. Specifically, we consider a compact star with a central energy density $\varepsilon(r=0) = \varepsilon_c$ and a total mass $M$, calculated using the boundary condition $P(R) = 0$, where $R$ is the radius of the star. Solving the TOV equation requires a particular EoS, which relates the pressure to the energy density within the star. 

\subsection{Tidal deformability}\label{ap}
The dimensionless tidal deformability can be defined as 
\begin{equation}
    \Lambda \equiv \frac{\lambda}{M^5} = \frac{2}{3} k_2 \left( \frac{R}{M} \right)^5,
    \label{eqq4}
\end{equation}
where $k_2$ is the quadrupole electric tidal Love number~\cite{Hinderer:2007mb} given by
\begin{align}
    k_2 &= \dfrac{8\tilde{C}^5}{5}(1-2\tilde{C})^2[2+2\tilde{C}(y_R-1)-y_R]\times\nonumber\\
    &\Big[ 2\tilde{C}[6-3y_R + 3\tilde{C}(5y_R-8)] +\nonumber\\
    &4\tilde{C}^3[13-11y_R+ \tilde{C}(3y_R- 2)+ 2\tilde{C}^2(1+y_R)] +\nonumber\\
    &3(1-2\tilde{C})^2[2-y_R + 2\tilde{C}(y_R - 1)]\ln(1-2\tilde{C})
    \Big]^{-1},
\end{align}
where $\tilde{C} = M/R$, is the compactness of the star, and $y_R = y(r=R)$, a dimensionless quantity associated with the internal solution of the associated perturbed metric~\cite{hinderer2010tidal}.
In this way, the Love number is a measurable quantity in gravitational wave signals that can be used to obtain
Information about the internal structure of a compact star. The expression for $k_2$ is valid for hadronic stars.  On the other hand, due to the self-bound of the SQM in forming QSs and the eminent discontinuity behavior at the surface of the star, $y_R$ is modified for QSs, such that
\begin{equation}
    y_R \rightarrow y_R - \dfrac{4\pi R^3{\Delta\varepsilon}}{M},
    \label{eqdelta}
\end{equation}
 where ${\Delta\varepsilon}$ denotes the difference in the energy density at the surface ($P=0$)  and the exterior ($\varepsilon=0$) of the QS~\cite{Lourenco:2021lpn, Postnikov:2010yn, Lopes:2021yga}. Thus, the EoSs that lead to similar mass-radius relations as the hadronic stars can present significantly different deformability if the QSs lead to greater or lesser energy density at the star's surface. This is reflected in Figs.~\ref{fig6} and~\ref{fig8} where almost all the mass-radius curves satisfy PSR J0030 + 0451 (presumably a hadronic star) with significantly larger $\Lambda_{1.4}$ values. In the EoSs for the MDDQM model, the value of  ${\Delta\varepsilon}$ is $\sim 102$MeV, while for a typical hadronic model where Baym-Pethick-Sutherland model (BPS)~\cite{Baym:1971pw} is used to simulate NS outer crust, the value is as low as ${\Delta\varepsilon} \sim 1.5 \times 10^{-8}$MeV, which is usually neglected in the determination of $\Lambda_{1.4}$. Clearly, for QSs $\Delta\varepsilon$ is too large to be ignored.

\bibliography{references.bib}

\end{document}